\documentclass[10pt,aps,prd,fleqn,superscriptaddress,notitlepage,nofootinbib,preprintnumbers,nobalancelastpage]{revtex4-1}
\PassOptionsToPackage{%
  breaklinks,
  colorlinks,
  urlcolor=blue,
  citecolor=citcolor,
  linkcolor=lcolor,
  linktoc=all
}{hyperref}
\usepackage[utf8]{inputenc}
\usepackage[T1]{fontenc}
\usepackage{graphicx}
\usepackage{dcolumn}
\usepackage{bm}
\usepackage{xspace}
\usepackage{amsmath}
\usepackage{amssymb}
\usepackage{mciteplus}
\usepackage{footnote}
\usepackage{feynmp-auto}
\usepackage{tikz}
\usepackage{orcidlink}
\usepackage{todonotes}
\definecolor{lcolor}{rgb}{0.5,0,0}
\definecolor{citcolor}{rgb}{0,0,1}
\usepackage{hyperref}
\usepackage{cleveref}
\usepackage{makecell}
\usepackage{booktabs,siunitx} 
\hypersetup{
    colorlinks=true,
    linkcolor=blue,
    citecolor=blue,
    filecolor=blue,
    urlcolor=blue
}
\newcommand{\sinSqW}{\ensuremath{\sin^2\theta_W}}
\newcommand{\LO}{\ensuremath{\text{LO}}\xspace}
\newcommand{\NLO}{\ensuremath{\text{NLO}}\xspace}
\newcommand{\NLOEW}{\ensuremath{\text{NLO}_\text{EW}}\xspace}
\newcommand{\NLOQED}{\ensuremath{\text{NLO}_\text{QED}}\xspace}
\newcommand{\NNLOQED}{\ensuremath{\text{NNLO}_\text{QED}}\xspace}
\newcommand{\csew}{\ensuremath{\text{CS}_{\mathbf{NLO_{EW}}}}\xspace}
\newcommand{\NNLOEW}{\ensuremath{\text{NNLO}_\text{EW}}\xspace}
\newcommand{\NNLO}{\ensuremath{\text{NNLO}}\xspace}

\newcommand{\Sherpa}{S\protect\scalebox{0.8}{HERPA}\xspace}
\newcommand{\Comix}{C\protect\scalebox{0.8}{OMIX}\xspace}

\newcommand{\Amegic}{A\protect\scalebox{0.8}{MEGIC}\xspace}

\newcommand{\betatilde}[2]{\ensuremath{\tilde{\beta}_{#1}^{#2}}\xspace}
\newcommand{\eik}[1]{\ensuremath{\tilde{S}\left(k_{#1}\right)}\xspace}

\newcommand{\ee}{\ensuremath{e^+e^-}\xspace}

\newcommand{\ddone}{{\mathrm d}}
\newcommand{\bea}{\begin{align}}
\newcommand{\eea}{\end{align}}
\newcommand{\beq}{\begin{equation}}
\newcommand{\eeq}{\end{equation}}

\newcommand{\OpenLoops}{O\protect\scalebox{0.8}{PEN}L\protect\scalebox{0.8}{OOPS}\xspace}
\newcommand{\Recola}{R\protect\scalebox{0.8}{ECOLA}\xspace}
\newcommand{\Griffin}{G\protect\scalebox{0.8}{RIFFIN}\xspace}   
\newcommand{\Muone}{M\protect\scalebox{0.8}{UONE}\xspace}   
\newcommand{\Moller}{M\protect\scalebox{0.8}{OLLER}\xspace}

\newcommand{\yfslo}{\ensuremath{\text{YFS}_{\mathbf{LO}}}\xspace}
\newcommand{\yfsnlo}{\ensuremath{\text{YFS}_{\mathbf{NLO_{EW}}}}\xspace}
\newcommand{\nloew}{\ensuremath{\text{YFS}_{\mathbf{NLO_{EW}}}}\xspace}

\newcommand{\yfsnnlo}{\ensuremath{\text{YFS}_{\mathbf{NNLO_{EW}}}}\xspace}
\newcommand{\UGeV}{\ensuremath{\mathrm{GeV}}\xspace}
\newcommand{\oforder}[1]{\ensuremath{\mathcal{O}\left(#1\right)}\xspace}
\newcommand{\MSbar}{\ensuremath{\overline{\mathrm{MS}}}\xspace}
\newcommand{\aew}{\ensuremath{\alpha_{\text{EW}}\xspace}}
\newcommand{\citenlotools}{\cite{Schonherr:2017qcj,Frederix:2018nkq,Bredt:2022dmm}\xspace}
\newcommand{\citenlosub}{\cite{Catani:1996vz,Frixione:1995ms,Gehrmann-DeRidder:2005btv,Boughezal:2011jf,Czakon:2010td,Catani:2002hc,Cacciari:2015jma,TorresBobadilla:2020ekr}\xspace}
\newcommand{\citenloslice}{\cite{Kunszt:1992tn,Fabricius:1981sx,Kramer:1986mc,Gaunt:2015pea,Giele:1991vf,Stewart:2010tn,Catani:2007vq}\xspace}
\newcommand{\citeoneloop}{\cite{Actis:2016mpe,Buccioni:2019sur,Berger:2009dq,Cullen:2011xs,Campbell:2010ff,vanHameren:2010cp,Denner:2014gla}\xspace}
\newcommand{\citetwoloop}{\cite{Freitas:2023iyx,Chen:2022dow,Freitas:2022hyp,Song:2021vru,Dubovyk:2018rlg,Delto:2023kqv,Bern:2000ie,Penin:2005eh,Becher:2007cu,
Penin:2005kf,Penin:2005eh,Becher:2007cu,Actis:2007gi,Penin:2016wiw,Bonciani:2004gi,Penin:2011aa,Czakon:2004wm,Bonciani:2021okt,Armadillo:2025mfx}\xspace}
\newcommand{\citeps}{\cite{Sjostrand:1985xi,Schumann:2007mg,Corcella:2000bw,
CarloniCalame:2000pz,Seymour:1991xa,CarloniCalame:2001ny,CarloniCalame:2005vc,
Barberio:1990ms,Arbuzov:2012dx,Masouminia:2021kne,Chen:2016wkt,Christiansen:2014kba,
Kleiss:2020rcg,Brooks:2021kji,Krauss:2014yaa,Christiansen:2015jpa,Dittmaier:2025htf}\xspace}
\newcommand{\citeLowEE}{\cite{Shatunov:2000zc,Aulchenko:2001je,Ignatov:2019omb,BaBar:2001yhh,BESIII:2009fln,Amelino-Camelia:2010cem,BaBar:2012bdw,BESIII:2015equ,KLOE:2004lnj,KLOE:2008fmq,KLOE:2010qei,CMD-2:2001ski}}
\def\emupm{\ensuremath{\mu^{\pm}e^-\rightarrow\mu^{\pm}e^-}}
\begin{document}

\preprint{IPPP/25/86, MCNET-25-28}
\title{\textbf{Towards a Fully Automated Differential $\NNLOEW$ Generator for Lepton Colliders} 
}%

\author{Alan~Price\orcidlink{https://orcid.org/0000-0002-0372-1060}}
\email{alan.price@uj.edu.pl}
\affiliation{Jagiellonian University, ul.\ prof.\ Stanis\l{}awa \L{}ojasiewicza 11, 30-348 Krak\'{o}w, Poland}

\author{Frank~Krauss\orcidlink{https://orcid.org/0000-0001-5043-3099}}
\email{frank.krauss@durham.ac.uk}
\affiliation{Institute for Particle Physics Phenomenology, Durham University, Durham DH1 3LE, UK}

\date{\today}

\begin{abstract}
\noindent
    Future proposed lepton collider experiments will reach unprecedented levels of accuracy. 
    To ensure the success of these experiments, and to fully exploit their wealth of data, the precision of theory calculations must reach comparable or even better levels. 
    One bottleneck in achieving this precision target lies in the systematic, process-independent inclusion of higher-order corrections at Next-to-Next-to-Leading Order in the electroweak coupling (\NNLOEW) while ensuring the correct matching  with modern all-orders resummation techniques. 
    Here, we present a solution to this problem, based on the Yennie-Frautschi-Suura theorem, which employs a local infrared (IR) subtraction to remove divergences and its matching to an all-order resummation of the soft and soft-collinear logarithms. 
\end{abstract}

\maketitle

\tableofcontents

\begin{center}
    \vspace*{3cm}
    {\large \itshape We dedicated this paper to the memory of Stanisław Jadach.}
\end{center}

\section{Introduction}
\noindent 
The current experiments under consideration for \ee colliders~\cite{ILC:2013jhg,Aicheler:2012bya,FCC:2018evy,CEPCStudyGroup:2018ghi,Vernieri:2022fae} offer an exceptionally promising environment for future discoveries, by improving our current bounds on new physics by several orders of magnitude.
As a result, these colliders are central to the future of particle physics and feature prominently in strategic planning efforts, including the European Strategy for Particle Physics~\cite{EuropeanStrategyforParticlePhysicsPreparatoryGroup:2019qin} and its ongoing update.
Central to this role is their ability to further probe the properties of the Higgs boson~\cite{:2012gk,Chatrchyan:2012ufa,ATLAS:2022vkf,CMS:2022dwd} and its role as a portal to potential new physics~\cite{Abramowicz:2016zbo,Robson:2018zje,DeBlas:2019qco,deBlas:2022ofj,deBlas:2019rxi,Artoisenet:2013puc,Barklow:2017suo,Gritsan:2022php,Ono:2012oyw,DiMicco:2019ngk}; they will also measure at unprecedented accuracy the electroweak pseudo-observables~\cite{ALEPH:2005ab}.
Especially for the latter, the accurate modelling of QED effects will prove hugely influential~\cite{Blondel:2018mad}.
This is just one example for how a successful physics program necessitates a significant improvement in theoretical modelling in parallel to the technical and experimental difficulties that need to be resolved. 
In particular, general purpose Monte-Carlo (MC) generators~\cite{Campbell:2022qmc} will need to undergo significant improvements~\cite{Freitas:2019bre,Heinemeyer:2021rgq} in their transition from providing high-quality simulations at the LHC, usually dominated by QCD effects, to playing a similarly pivotal role at lepton colliders, where typically electroweak corrections are most significant.
In the past years significant progress has been made towards the automation of next-to-leading order (\NLO) calculations for \ee~\citenlotools, as well as further developments in process specific generators~\cite{Jadach:2022mbe,CarloniCalame:2019nra,Banerjee:2020rww}.
To ensure our theoretical predictions can match the precision of future experiments we need to push the precision frontier in which the inclusion of higher-order, at least up to \NNLOEW, will become mandatory. 
However, simply including fixed-order predictions will not be sufficient without a comparable matching to a resummation scheme that will account for potentially large logarithms that could spoil their parametric accuracy. 
In full analogy to similar calculations for the LHC, this can be achieved in the collinear picture by convoluting the electron parton distribution function (PDF)~\cite{Skrzypek:1990qs}, known to \NLO~\cite{Bertone:2019hks} and recently to \NNLO~\cite{Stahlhofen:2025hqd,Schnubel:2025ejl}, with a short distance partonic cross-section that can be calculated to higher-orders with standard techniques~\cite{Frederix:2009yq,Catani:1996vz}.
Results of such calculations can be combined with a parton  shower~\citeps to achieve fully exclusive predictions~\cite{Frixione:2007vw,Frixione:2002ik}.
While there has been significant progress in the inclusive picture~\cite{Blumlein:2020jrf,Blumlein:2019srk} the implementation of a fully matched \NLOEW/\NNLOEW parton shower for arbitrary processes is so far missing.

\noindent
An alternative approach to collinear resummation is available through the Yennie-Frautschi-Suura (YFS) theorem~\cite{Yennie:1961ad}, in which the entire perturbative expansion is reordered such that the infrared (IR) divergences, arising from soft photon emissions, are subtracted order-by-order while the, potentially large, associated logarithms are resummed to infinite order. 
This reordering defines an all-order subtraction scheme that can be used to calculate higher-order corrections. 
In addition, the YFS theorem can be cast into a MC algorithm~\cite{Jadach:1988gb,Krauss:2022ajk} that treats the complete photon phasespace analytically, allowing the MC to create explicit photon kinematics. 
As a result, once the YFS calculation is accurate to a given order, the matching of the multi-photon emissions is achieved automatically. 
To date this has only been achieve in process dependent codes based on the Coherent exclusive resummation (CEEX)~\cite{Jadach:2000ir}, where 
the IR subtraction is performed at the amplitude level.
In this paper, we will show how to construct the finite remainders of the YFS theorem to \NNLOEW in an automated way at the level of squared amplitudes. 
We have implemented this method in the \Sherpa~\cite{Bothmann:2019yzt,Sherpa:2024mfk} event generator, where the matching is fully automated at \NLOEW (and called \yfsnlo in the following). 
For the matching to \NNLOEW, we have implemented the double-real and real-virtual subtractions in an automated fashion.
We have also implemented the subtraction term for the double-virtual contribution, but due to a lack of a publicly available tool for the automatic calculation of two-loop EW amplitudes we are restricted to simple $2\rightarrow2$ processes where results can be found in the literature, cf.~\cite{Berends:1987ab,Chen:2022dow}.
We stress that it will be a straightforward exercise to promote \Sherpa to complete \yfsnnlo accuracy once such an automated tool becomes available.
 
\section{Theory}

\noindent 
The YFS theorem facilitates the resummation, to all orders, of logarithms associated with the emission of multiple, and in principle infinitely many, soft and soft-collinear photons, emitted by massive particles. 
The theorem also provides a complete analytical treatment of the multi-photon phasespace in which the photons are {\em explicitly} generated, with a resolution criterion given by a combined energy and angular cut-off. 
Reconstructing the full kinematic structure of scattering events leads to a straightforward implementation of the YFS method as both a cross-section calculator and an event generator. 
While both the soft and soft-collinear logarithms are resummed, the remaining collinear logarithms, of the form $\alpha^m \log^n\left(Q^2/m_\ell^2\right)$, can be introduced in a systematic manner order-by-order in \aew, while divergences associated with such emissions are regulated by the non-zero fermion masses. 
With the IR divergences extracted and resummed to all orders, the differential cross-section can be expressed as, 
\begin{align}\label{eq:masterYFS}
    &\ddone\sigma^{\left(\infty\right)} = \sum_{n_\gamma=0}^\infty
        \frac{e^{Y(\Omega)}}{n_\gamma!}\,
        \ddone{\Phi_n} 
        \left[\prod_{i=1}^{n_\gamma}\ddone{\Phi_i^\gamma}\,\eik{i}\,\Theta(k_i,\Omega)\right]
        \left(\tilde{\beta}_0\left(\Phi_n\right)
        + \sum_{j=1}^{n_\gamma}\frac{\tilde{\beta}_{1}(\Phi_{n+1})}{\eik{j}}
        +\sum_{{j,k=1}\atop{j< k}}^{n_\gamma}
        \frac{\tilde{\beta}_{2}(\Phi_{n+2})}{\eik{j}\eik{k}} 
        + \cdots\right)\,,
\end{align}
where $\eik{i}$ is is the eikonal factor for a given (charged) dipole $ij$ emitting a photon $k$. 
It is given by,
\begin{align}\label{eq:eik}
\tilde{S}\left(k\right) = \sum_{lm} \tilde{S}_{lm}\left(k\right),\,\,\,\,
\tilde{S}_{lm}\left(k\right) = \left(j^\mu_l\left(k\right) + j^\mu_{m}\left(k\right)\right)^2,\,\,\,\,
j^\mu_l\left(k\right) = Z_lQ_l\frac{p_l^\mu}{p_lk}
\end{align}
where the sum runs over all dipoles ($lm$), $Q_l$ is the QED charge and $Z_l=1(-1)$ for incoming (outgoing) particles.
$Y(\Omega)$ is the YFS form-factor~\cite{Jadach:2000ir,Jadach:1996hi,Schonherr:2008av},
\begin{equation}\label{EQ::FormFactor}
Y(\Omega) =
\sum_{i<j}\mathcal{R}e\;\mathcal{B}_{ij}(\Phi_n) +
    \tilde{\mathcal{B}}_{ij}(\Phi_{n+1})\,,
\end{equation} 
where $\Omega$ represents the soft photon domain.
In the above expression we have suppressed the indexing of unresolved emissions by introducing the following notation,
\begin{align}
\tilde{\beta}_{n_R} = \sum_{n_V=0}^\infty\tilde{\beta}_{n_R}^{n_V+n_R}\,,
\end{align} 
where $n_R$, $n_V$ are the number of resolved and unresolved ({\it i.e.}\ virtual) emissions under consideration. 
The first term in~\cref{eq:masterYFS} that contains the multiple photon phasespace measure can be used to derive an algorithm for the generation of the complete photon phasespace in an MC generator.
It can be implemented in a process-independent manner, see, e.g.~\cite{Krauss:2022ajk}. 
The remaining terms of~\cref{eq:masterYFS}, within the brackets, represent the higher-order residuals which remain after the resummation of the IR-divergent terms. 
These are the expressions that can be used to recursively implement the systematic matching of higher-order corrections. 

\noindent
When including higher-order corrections it is crucial that all IR singularities are regulated, either through a suitable subtraction scheme~\citenlosub or by phase-space slicing methods~\citenloslice.
The YFS theorem can be categorized as the former, as it facilitates the construction of local subtraction terms that cancel the corresponding IR singularities. 
A unique feature of the YFS subtraction scheme is that it does not require the introduction of additional terms to ensure local IR finiteness.
In general, subtraction schemes guarantee overall IR finites by virtue of the Kinoshita–Lee–Nauenberg (KLN) theorem~\cite{Kinoshita:1962ur,Lee:1964is}, in which the combination of unresolved and resolved contributions ensure that the IR singularities cancel. 
In practice, these IR poles appear in separate independent integrals, the former in a $n$-body phasespace and the latter in a $n+1$-body phasespace. 
In most cases both must be treated numerically. 
This typically requires the introduction of additional subtraction terms within each phasespace, which render each term locally IR finite while leaving the total contribution unchanged.
In the YFS scheme, the IR poles are not cancelled according to the KLN theorem but rather each individual residue in~\cref{eq:masterYFS} is by itself IR finite. 
Essentially, we can take any IR divergent correction of order \oforder{\alpha^{n_V+n_R}}, where $n_V=n_R$ is not required, and construct local subtraction terms. 
The only constraint for constructing such terms is that all corrections that are of a lower order must be completely constructed as they will enter the expressions for the local subtraction terms. 
In practice, it is not possible to construct an all-order matching: while the structure of the subtraction terms is known to all-orders, we are limited by our ability to calculate the amplitudes themselves. 
At the moment, the calculation of amplitudes at \NLOEW has essentially been automatised for arbitrary processes at both lepton and hadron colliders, taking advantage in particular of the advent of automatic tools~\citeoneloop for the evaluation of one-loop diagrams and the implementation of automatic subtraction schemes in the MC generators~\citenlotools. 
As we will show in the next sections we have automatised the complete YFS subtraction at \NLOEW for arbitrary processes at lepton colliders. 
For \NNLOEW, it is the contribution of the two-loop diagrams which is causing the bottleneck for our automation.
While it is possible to get the double-real and real-virtual amplitudes, as the former is tree-level and the latter is a one-loop contribution, both of which can be calculated using the standard tools, there is currently no public automatic tool for the evaluation of two-loop amplitudes for \NNLOEW
\footnote{However it is possible to include some universal corrections that have been previously implemented in \Sherpa ~\cite{Krauss:2022ajk}
which could be of interest in future studies.}. 

\noindent
Originally, the YFS corrections employed within \Sherpa were limited to the leptonic decays of bosons~\cite{Schonherr:2008av}, including exact \NLOQED correction for certain decay channels, in particular  $\tau\to\ell\nu_\ell\nu_\tau$ and some hadronic decays~\cite{Alioli:2016fum,Bernlochner:2010fc}. 
It was further extended in~\cite{Krauss:2018djz} to include \NLOQED $+$ \NLOEW and \NNLOQED $+$ \NLOEW corrections for $h\to\ell\ell$ and $Z\to\ell\ell$, while for the decay of charged bosons, $W\to\ell\nu$, the precision is limited to \NLOEW as the purely QED corrections cannot be defined in a gauge-invariant manner. 
This soft-limit based treatment of final state radiation (FSR) has been further augmented by universal corrections due to the emission of a hard collinear photon in~\cite{Flower:2022iew}. 
The YFS treatment in \Sherpa was further extended to include emissions from both the initial and final states in process-independent manner in~\cite{Krauss:2022ajk}. 
While the resummation has been automated, only process-specific perturbative corrections for some relatively simple processes were included, and they were limited to QED corrections only. 
In the next sections we will show how this initial treatment can be extended to a process-independent calculational setup, thereby increasing the theoretical and practical reach of our event generator.

\subsection{\protect\NLOEW Matching}\label{SubSec:NLO}
\subsubsection{Virtual terms}\label{SubSubSec:Virtual}
\noindent
To see explicitly how the subtraction scheme works at \NLOEW, let us isolate the IR finite residual for the unresolved emissions in~\cref{eq:masterYFS}.
It is given by,
\begin{equation}\label{EQ:OneLoopIR}
    \betatilde{0}{1}\left(\Phi_n\right) = \mathcal{V}(\Phi_n) - \sum_{ij}\mathcal{D}_{ij}\left(\Phi_{ij}\otimes\Phi_n\right).
\end{equation} 
In~\cref{EQ:OneLoopIR}, the first term represents the full one-loop electroweak correction to an arbitrary process at leading order (Born level, LO), which will typically contain IR divergences, and the second represents the corresponding YFS subtraction term. 
The one-loop contribution, $\mathcal{V}(\Phi_n)$, can be computed using standard tools such as \OpenLoops~\cite{Buccioni:2019sur} or \Recola~\cite{Actis:2016mpe,Denner:2017wsf} which have been interfaced to \Sherpa~\cite{Biedermann:2017yoi}.
They automatically provide loop-level amplitudes for arbitrary SM processes. 
The subtraction term is constructed, for a given phasespace point, as a sum over all contributing charged dipoles, $ij$, and for each dipole we apply the YFS resummation algorithm. 
Carefully avoiding double counting allows us to construct the subtraction term from the Born contribution and the virtual form-factor,
\begin{equation}
\mathcal{D}_{ij}\left(\Phi_{ij}\otimes\Phi_n\right) = \betatilde{0}{0}\left(\Phi_n\right) \mathcal{R}e\;\mathcal{B}_{ij}(\Phi_n).
\end{equation} 
The IR structure of both these terms can be extracted either by using dimensional regularization~\cite{tHooft:1972tcz} (DR), in which both terms are evaluated in $d=4-2\epsilon$ dimensions, or by introducing a fictitious photon mass $M_\gamma$ with which we regularize the IR divergence. 
With massive regularization, the IR divergences manifest as logarithms of $M_\gamma$ while in DR they will be represented by poles of the form $\epsilon^{-1}$.
Both approaches have been implemented in \Sherpa allowing us to perform dedicated cross-checks between the two schemes.
At \NLOEW we can directly convert between the two by identifying,
\begin{equation}
\ln\left(M_\gamma^2\right) = \frac{(4\pi\mu^2)^\epsilon}{\epsilon\Gamma(1-\epsilon)}+\oforder{\epsilon} = -\frac{1}{\epsilon}+\log\left(\frac{4\pi\mu^2}{\Delta^2}\right)+\oforder{\epsilon}
\end{equation}
where $\mu$ is the scale of DR and $\Delta^2$ is a scheme-dependent parameter which in the \MSbar scheme is simply $4\pi$.
\begin{figure}[!h]
    \centering
    \includegraphics[width=0.4\textwidth]{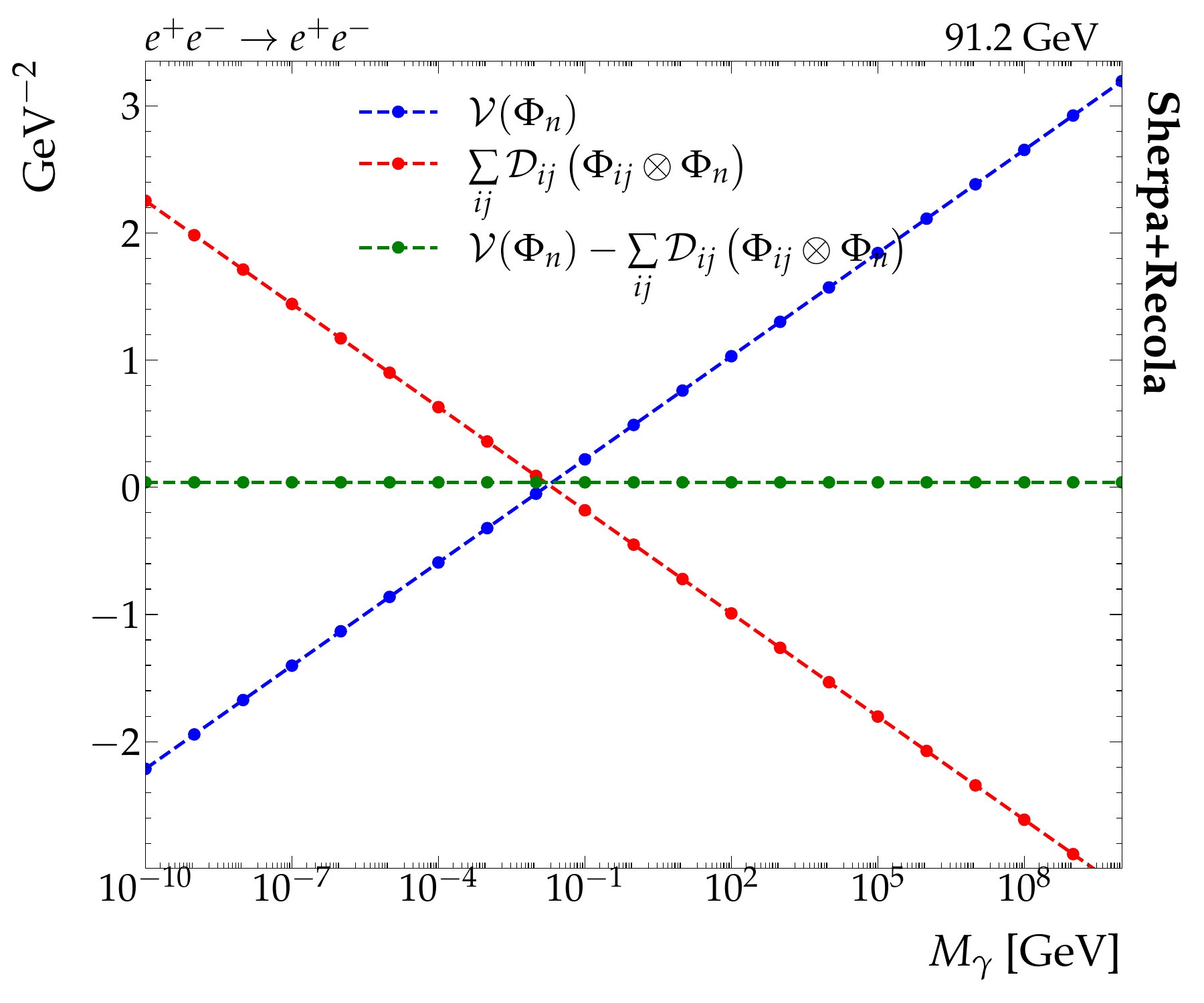}
    \includegraphics[width=0.4\textwidth]{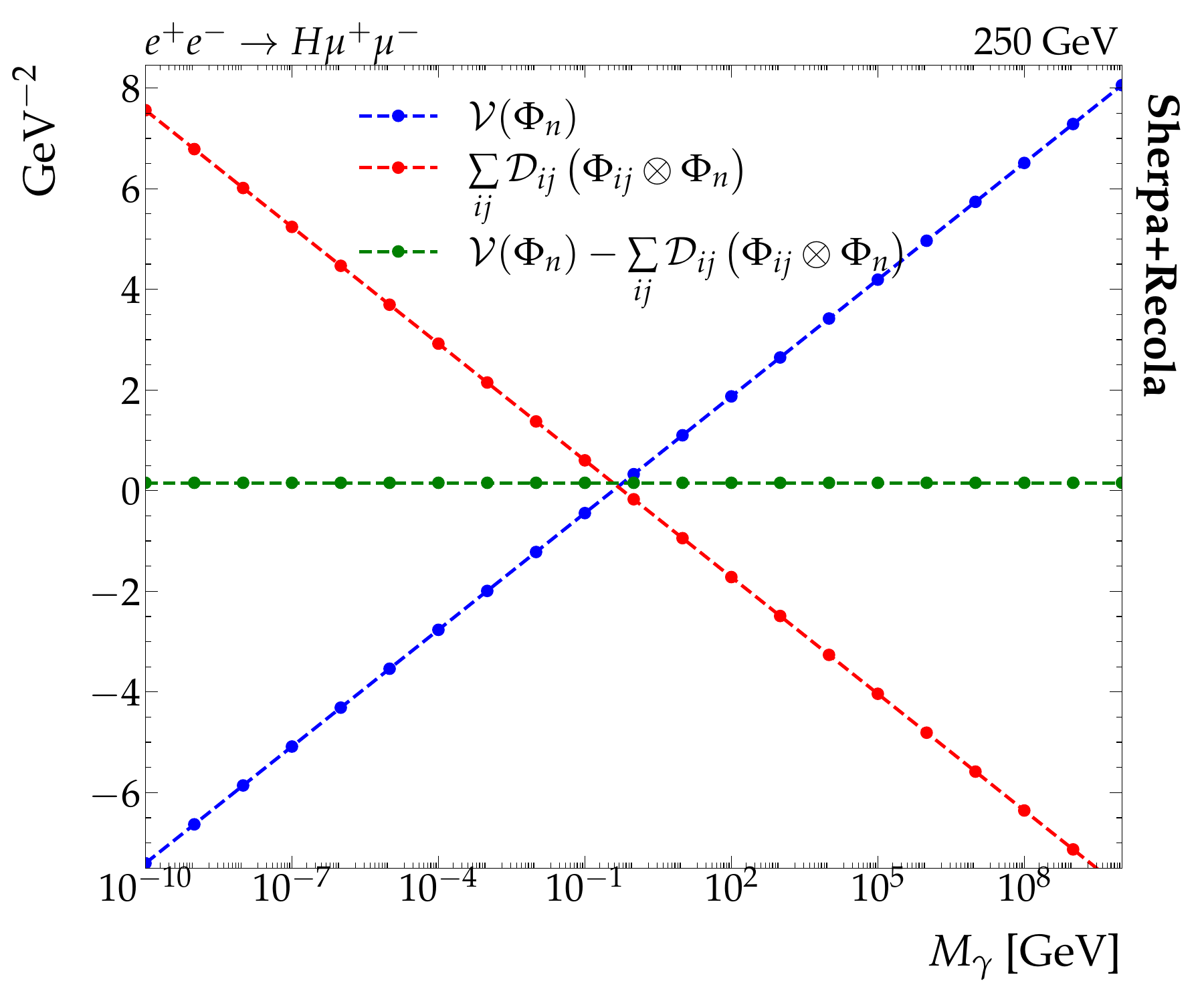}
    \includegraphics[width=0.4\textwidth]{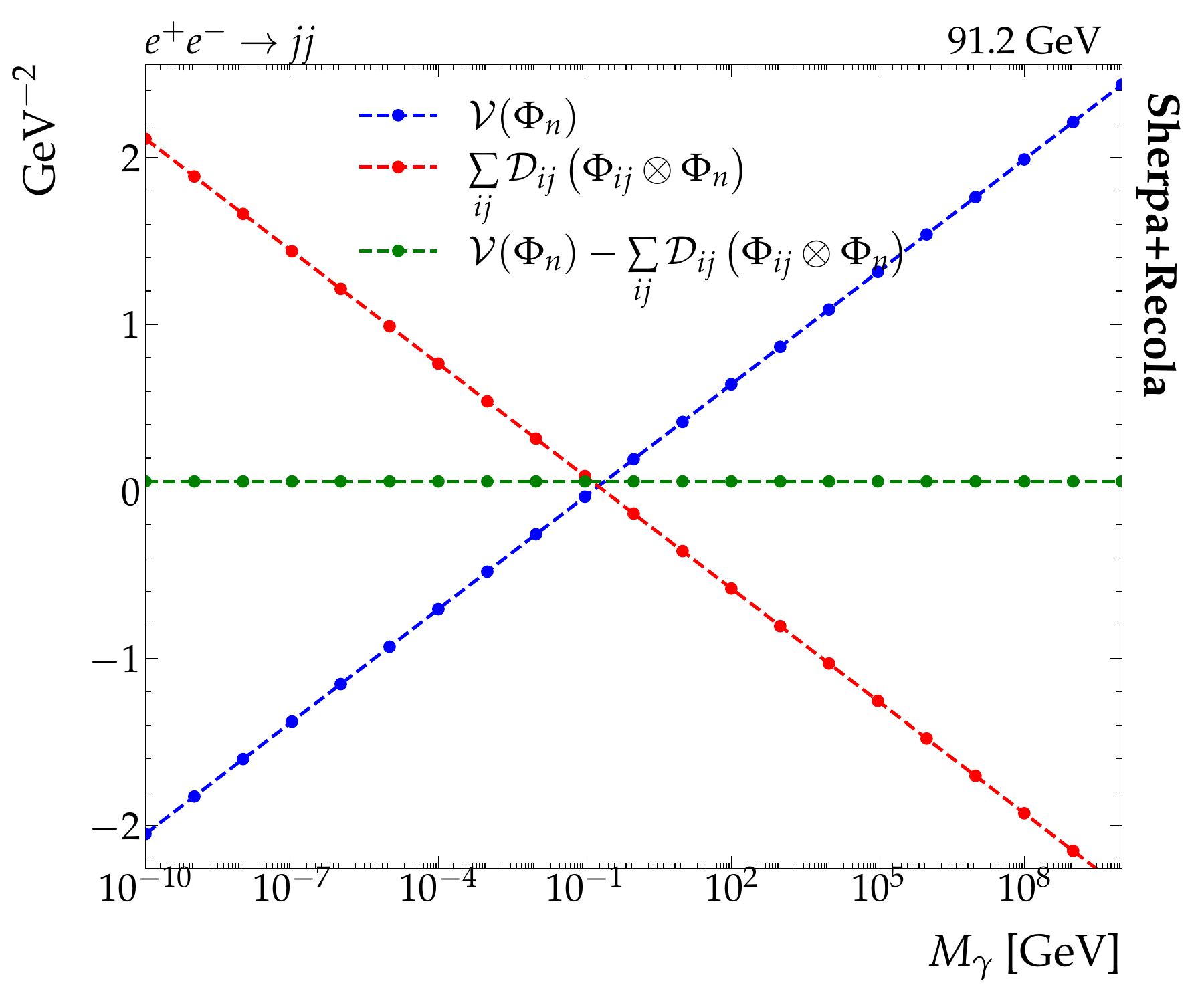}
    \includegraphics[width=0.4\textwidth]{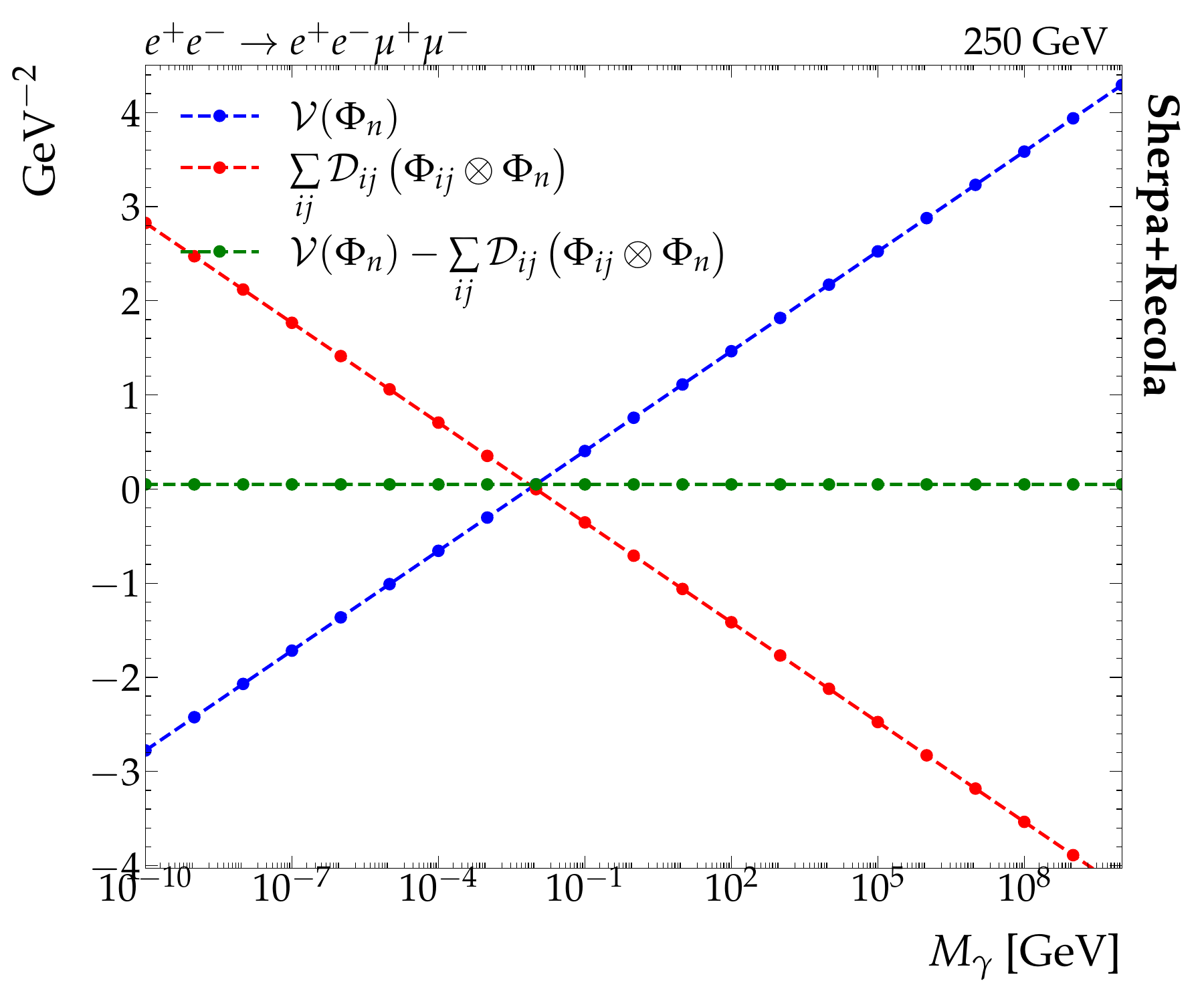}
    \caption{Explicit cancellation of infrared (IR) divergences, as described in~\cref{EQ:OneLoopIR}, using massive regularisation. 
    The plot includes the finite sum (green), as well as the individual contributions from the virtual term (blue) and the subtraction term (red). 
    The y-axes have been rescaled to improve readability.}
    \label{fig:massreg}
\end{figure}

\noindent 
In~\cref{fig:massreg}, we show both the complete IR finite result,~\cref{EQ:OneLoopIR}, as well as the individual components which we evaluate using a wide range of values for $M_\gamma$. 
We see in both the virtual and subtraction terms the expected logarithmic dependence on $M_\gamma$, while the sum of both terms does not depend on our choice. 
The green line, which represents the non-zero complete IR-finite residual, will be included in~\cref{eq:masterYFS}.

\begin{figure}[!h]
    \centering
    \includegraphics[width=0.4\textwidth]{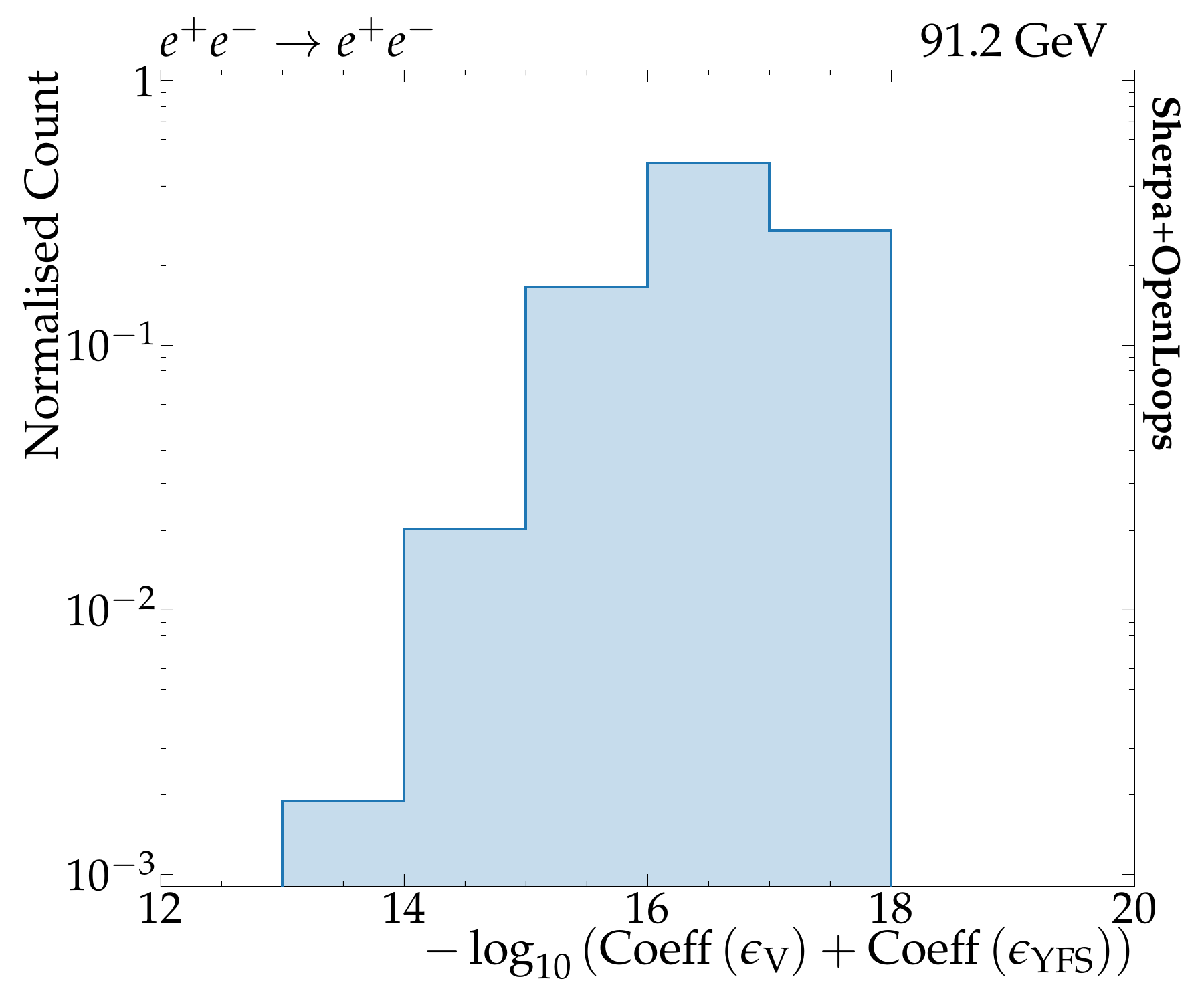}
    \includegraphics[width=0.4\textwidth]{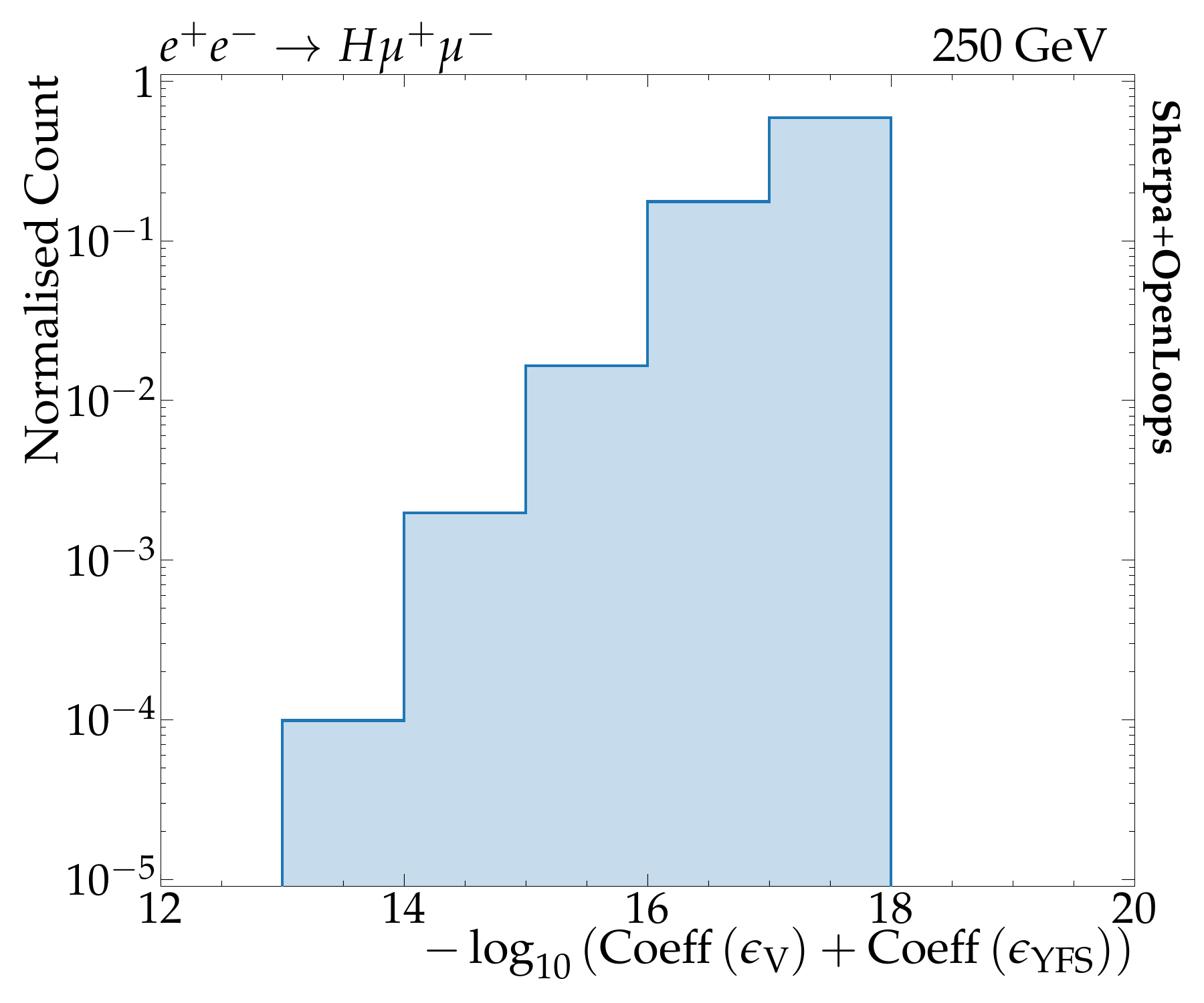}
    \includegraphics[width=0.4\textwidth]{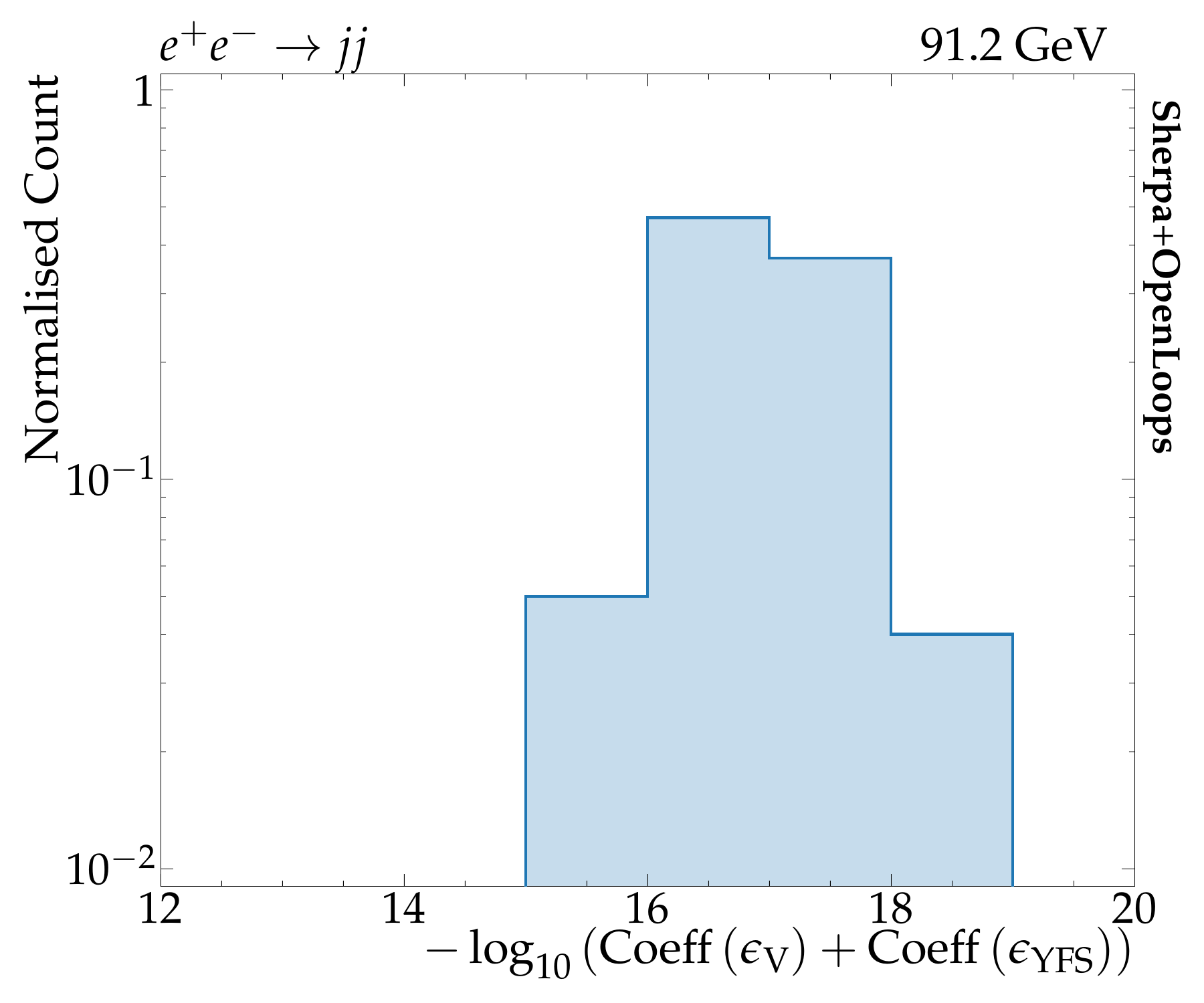}
    \includegraphics[width=0.4\textwidth]{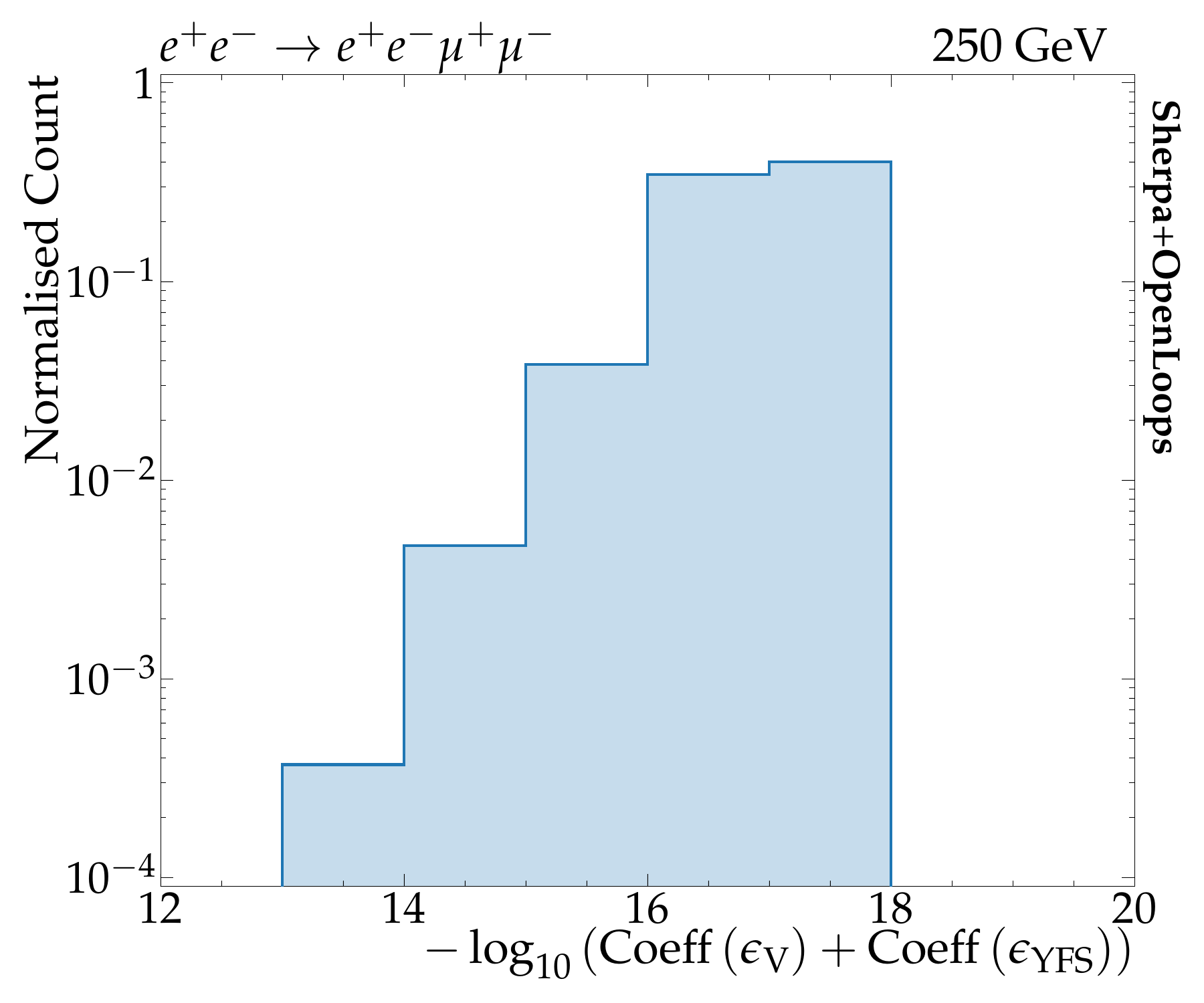}

    \caption{Cancellation of the IR poles according to ~\cref{EQ:OneLoopIR} using dimensional regularization. 
    The bins correspond to the number of decimal places at which the cancellation was achieved, typically at the level of 
    machine (double floating point) precision.
    Note that for processes which contain, e.g., 4 charged fermions in the final state in addition to the two incoming charged leptons, we are left with a total of 15 unique dipoles entering the subtraction. }
    \label{fig:DimReg}
\end{figure}
\noindent 
When implementing the subtraction in dimensional regularization one key requirement is the cancellation of the $\epsilon^{-1}$ poles between the virtual and the subtraction terms for each phasespace point. 
This check is performed internally within \Sherpa, where we compare our coefficient arising from the subtraction term with that of the one-loop providers. 
We summarize the agreement in~\cref{fig:DimReg}, where we exhibit the number of digits to which the two numbers agree. 
We find that in the majority of events, the two agree to 13-18 digits depending on the process in question, with the majority cancelling to at least 16 digits. 
This is within a satisfactory range of agreement within \Sherpa which currently does not support quad-level precision, which would allow us to probe the agreement with even higher accuracy. 
In addition in the YFS scheme we must keep all masses, and in particular the electron mass explicitly different from zero; this may lead to some smaller numerical instabilities impacting on the distribution of correct digits. 
In both analyses we have considered four separate final states for each method to highlight the automatic nature of the subtraction.

\subsubsection{Real terms}\label{SubSubSec:Real}
\noindent
For corrections arising from resolved emissions of photons we can also define an IR subtraction term within the YFS framework as,
\begin{align}\label{eq:RealSub}
\tilde{\beta}_1^1\left(\Phi_{n+1}\right) &= \mathcal{R}\left(\Phi_{n+1}\right) - \sum_{ij}\mathcal{\tilde{D}}_{ij}\left(\Phi_{ij+1}\otimes\Phi_n\right)\,,\\
 \mathcal{R}\left(\Phi_{n+1}\right) &=\frac{1}{2(2\pi)^3}\left|\mathcal{M}^{\frac{1}{2}}_1\left(\Phi_{n+1}\right)\right|^2
\end{align} 
where $\mathcal{M}^{\frac{1}{2}}_1\left(\Phi_{n+1}\right)$ is the complete $\mathcal{O}(\alpha)$ real correction. 
This amplitude, while containing potentially divergent terms, is a pure tree-level amplitude which can be calculated automatically using modern amplitude methods and tools~\cite{Krauss:2001iv,Gleisberg:2008fv,Mangano:2002ea,Kilian:2007gr,Alwall:2011uj}.~\footnote{ 
  In \Sherpa, there are two independent automated matrix-element generators (MEGs), namely \Amegic~\cite{Krauss:2001iv} and \Comix~\cite{Gleisberg:2008fv}, which have already been used in the automated calculation of the $\tilde{\beta}_0^0 \left(\Phi_{n}\right)$ terms in \Sherpa's original implementation of the YFS scheme.} 
The second term in~\cref{eq:RealSub} represents the YFS dipole subtraction term which, similarly to the virtual case, is constructed automatically within \Sherpa. 
It can be directly taken from the YFS form factor in analytical form~\cite{Jadach:1995sp} for a given charged dipole as,
\begin{equation}
\mathcal{\tilde{D}}_{ij}\left(\Phi_{ij+1}\otimes\Phi_n\right) = \tilde{S}_{ij}\left(k\right)\betatilde{0}{0}\left(\Phi_n\right)
\end{equation}
where $\betatilde{0}{0}$ is the LO squared amplitude and $\tilde{S}_{ij}\left(k\right)$ is the eikonal term, as defined in~\cref{eq:eik}.
\begin{figure}
    \centering
     \includegraphics[width=0.4\textwidth]{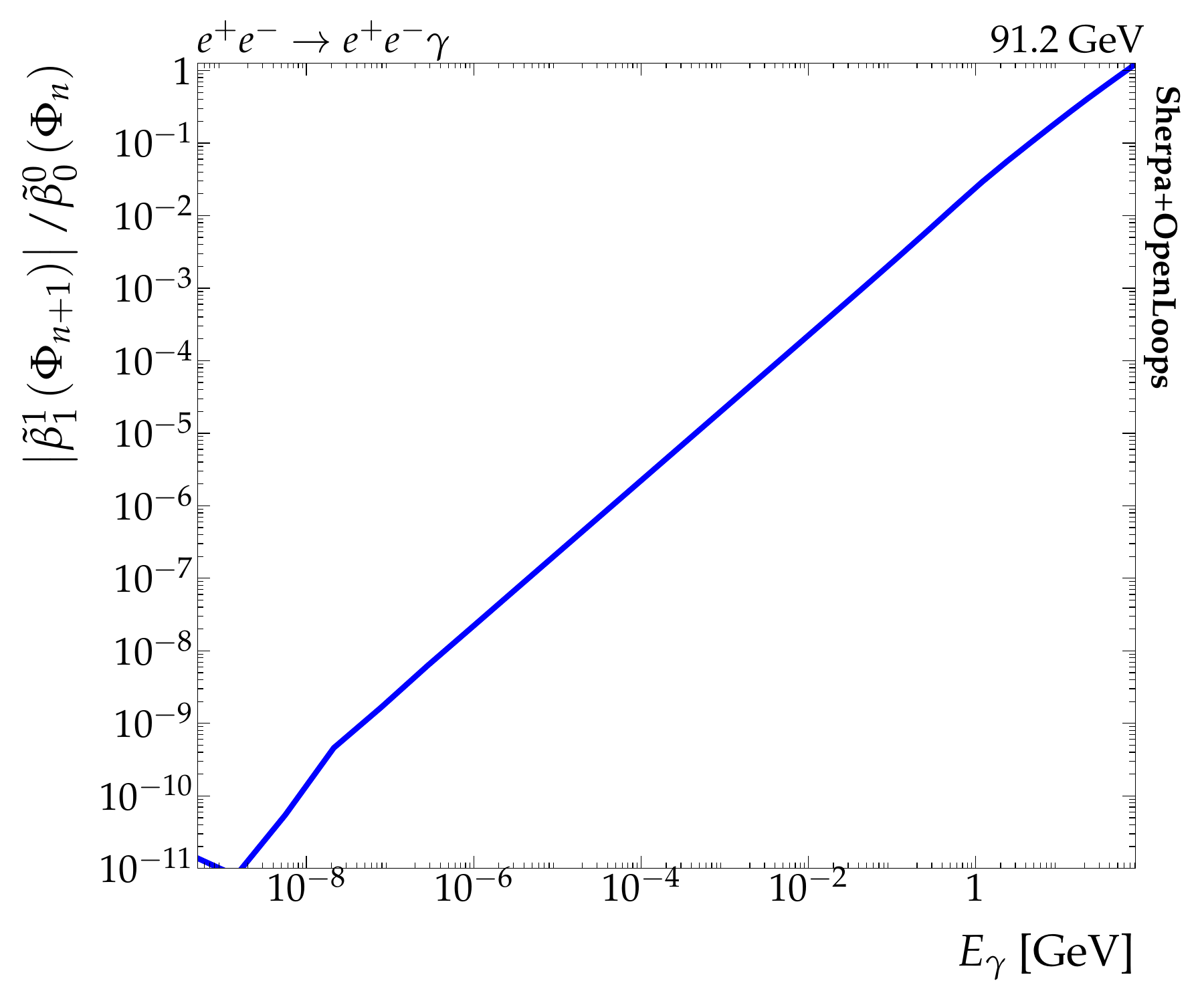}
    \includegraphics[width=0.4\textwidth]{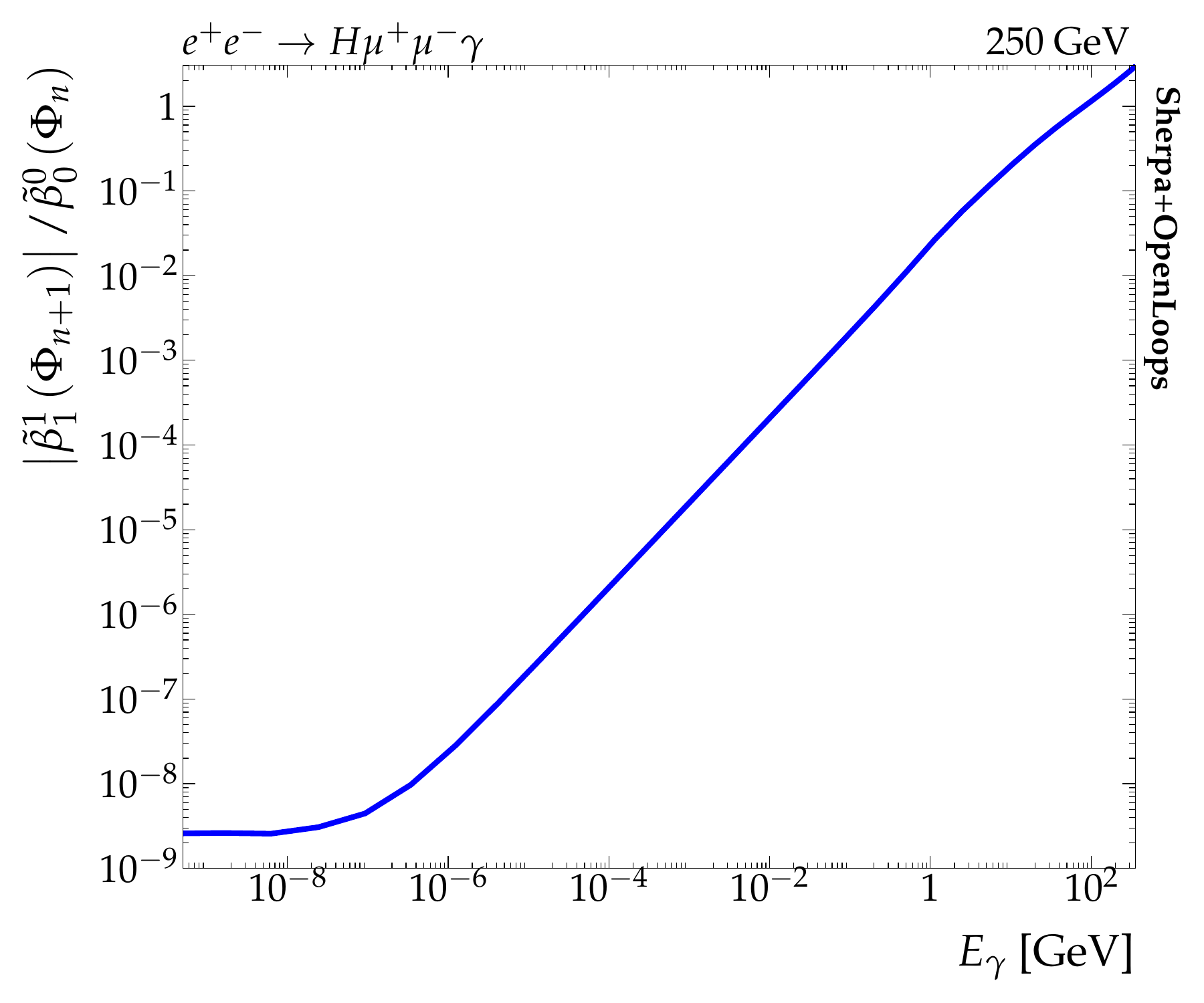}
    \includegraphics[width=0.4\textwidth]{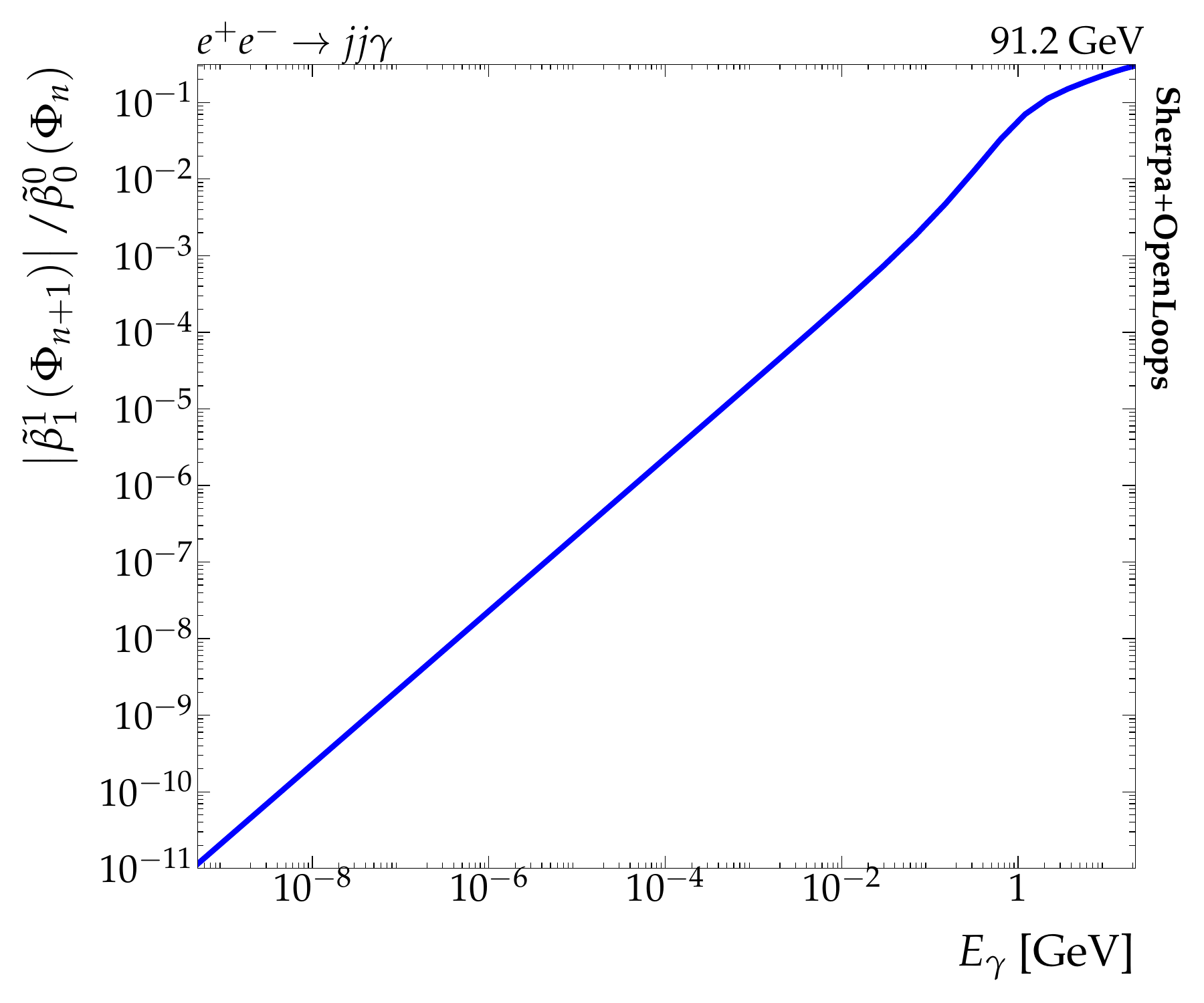}
    \includegraphics[width=0.4\textwidth]{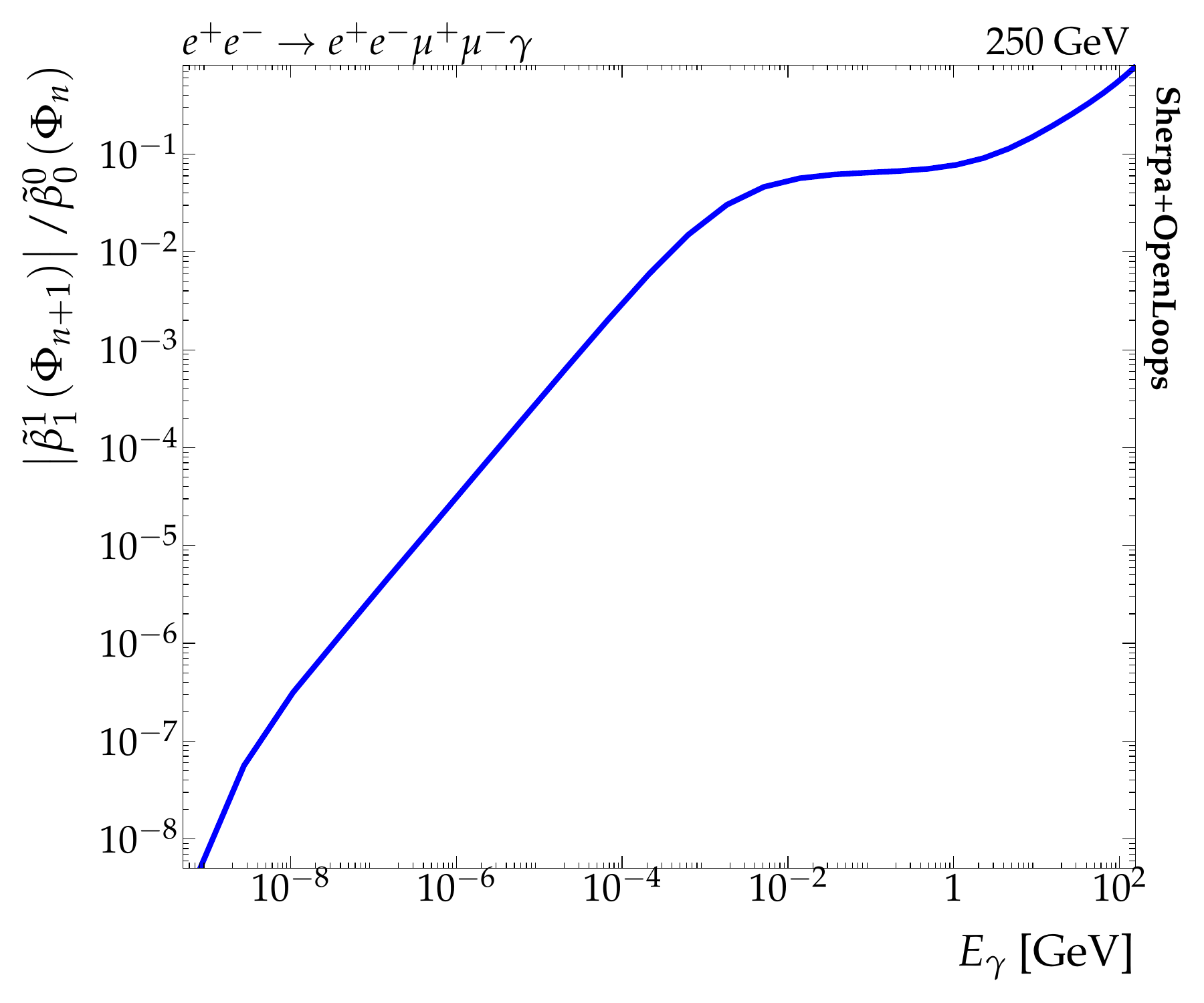}
    \caption{Magnitude of the subtracted real correction, normalized to the Born-level baseline, in the IR limit of vanishing photon momentum.}
    \label{fig:RealSub}
\end{figure}

\noindent
When evaluating~\cref{eq:RealSub} we face the fact that we are calculating an amplitude in $\Phi_{n+1}$ phasespace, which corresponds to the underlying Born process plus one photon.
However, since we are completely differential in the multi-photon phasespace we have to ensure that we have a proper mapping from the $\Phi_{n+n_R}$ to the $\Phi_{n+1}$ phasespace.  
This is achieved by realizing that~\cref{eq:RealSub} is derived exactly in the limit where all other photon momenta vanish. 
It is then sufficient to proceed with the original YFS phasespace mappings as if there was only one photon emission.
This argument holds to all orders for any given $n_R$ real emissions and can equally be applied to corrections involving additional virtual emissions. 
We shall discuss the explicit details of the mappings in~\cref{SubSec:MMaps}.

\noindent
Taking a closer look at the subtraction term for~\cref{eq:RealSub}, it is clear that in the extreme soft limit purely fixed-ordered predictions without any resummation will suffer from numerical instabilities since $\mathcal{R}\left(\Phi_{n+1}\right)$ will grow with inverse powers of the photon energy. 
In the same limit, in the YFS approach~\cref{eq:RealSub} becomes vanishingly small, a reflection of the fact that the IR behaviour of the real amplitude behaves as,
\begin{equation}
\lim_{k\rightarrow 0} \mathcal{R}\left(\Phi_{n+1}\right) \rightarrow   
2(2\pi)^3\,\sum\limits_{ij}\tilde{S}_{ij}\left(k\right)\betatilde{0}{0}\left(\Phi_n\right)\,,
\end{equation}
reproducing the subtraction term increasingly precisely.
While the subtracted term therefore vanishes in this limit we stress that the physical effects of soft photons have already been resummed into the YFS form factor and are accounted for. 
In~\cref{fig:RealSub}, we explore this limit for our example processes. 
For this test, we choose a phasespace point that contains at least one sufficiently hard photon and we evaluate~\cref{eq:RealSub}. 
We then artificially reduce the photon's energy, reapply the YFS phasespace mapping and again evaluate~\cref{eq:RealSub}~\footnote{ 
We continue this reduction procedure down until we reach the technical IR cut-off, the value of which we are completely 
free to chose, as below this cut-off the algorithm will fail and we can no longer guarantee IR finiteness. 
}
We use \OpenLoops to evaluate the real emission amplitudes as its internal numerical routines are highly stable in the ultra-soft region.
We recognise the same limiting behaviour, namely that the subtracted real contribution of~\cref{eq:RealSub} becomes vanishingly small in the soft limit, as expected from the YFS theorem. 
From these plots we deduce that the subtraction is stable for energies down to $10^{-9}$~\UGeV. 
The combination of these two corrections, ~\cref{eq:RealSub,EQ:OneLoopIR}, and the process independent implementation in \Sherpa allows the matching of our YFS resummation to the complete \NLOEW corrections.
The results of this procedure will be dubbed \yfsnlo in the following.

\subsection{\NNLOEW Matching}\label{SubSec:NNLO}
\subsubsection{Real-Virtual Correction}\label{SubSubSec:RV}
\noindent
The first \NNLOEW corrections that we will consider are the real-virtual emissions, essentially the one-loop corrections to the LO process plus an additional resolved photon emission. 
From the YFS theorem, we can extract these IR finite contributions as,
\begin{align}\label{EQ:RealLoopIR}
    \betatilde{1}{2}\left(\Phi_{n+1}\right) &= \mathcal{RV}(\Phi_{n+1}) -  \sum_{ij}\mathcal{D}^{(1)}_{ij}\left(\Phi_{ij+1}\otimes\Phi_n\right)\,,\\
    \mathcal{RV}(\Phi_{n+1}) &= \frac{1}{2(2\pi)^3}\left|\mathcal{M}^{\frac{3}{2}}_1\left(\Phi_{n+1}\right)\right|^2,
\end{align}
where $\left|\mathcal{M}^{\frac{3}{2}}_1\left(\Phi_{n+1}\right)\right|^2$ is the full real-virtual squared amplitude, constructed by combining tree-level and one-loop amplitudes as before. 
The dipole subtraction term can be constructed similarly to the real emission case, by multiplying the IR-finite virtual correction, \cref{EQ:OneLoopIR}, with the eikonal factor, \cref{eq:eik}, yielding,
\begin{equation}\label{EQ:RVSub}
\mathcal{D}^{(1)}_{ij}\left(\Phi_{ij+1}\otimes\Phi_n\right) = \tilde{S}_{ij}\left(k\right)\betatilde{0}{1}\left(\Phi_n\right)\,.
\end{equation}
Again, the summation runs over all possible combinations of charged particles; the added index reflects that this subtraction involves an unresolved emission. 
In~\cref{EQ:RVSub}, we see for the first time how the recurrence nature of the YFS subtraction manifests itself at \NNLO, as this subtraction term includes the previously calculated \NLOEW-finite contribution $\betatilde{0}{1}\left(\Phi_n\right)$. 
The $n+1$ loop amplitude can contain two types of IR singularities. 
The first one will manifest itself as $\epsilon$ poles in the one-loop amplitude, and the second type will arise when the resolved photon becomes soft. 
Both of these divergences can be removed by the local subtraction term in~\cref{EQ:RVSub}.
As this correction involves the emission of a resolved photon, we must also ensure that we correctly map the phasespace $\Phi_{n+n_k} \rightarrow \Phi_{n+1}$ as we did for the real emission case and ensure that we evaluate both terms in~\cref{EQ:RealLoopIR} within this reduced phasespace. 
In particular, when constructing the subtraction term, the eikonal is calculated using the local $\Phi_{n+1}$ phasespace point.

\begin{figure}
    \centering
    \includegraphics[width=0.4\textwidth]{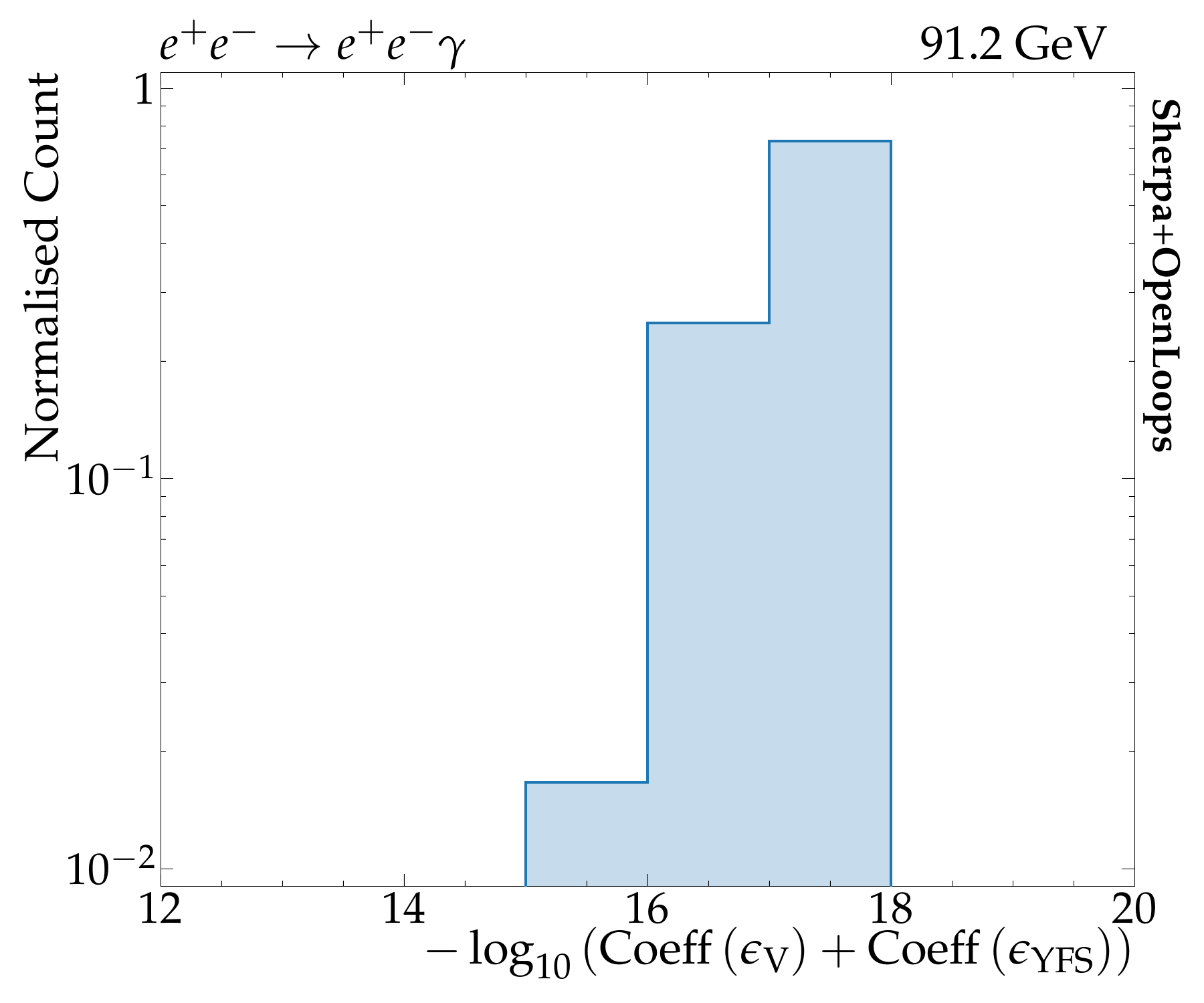}
    \includegraphics[width=0.4\textwidth]{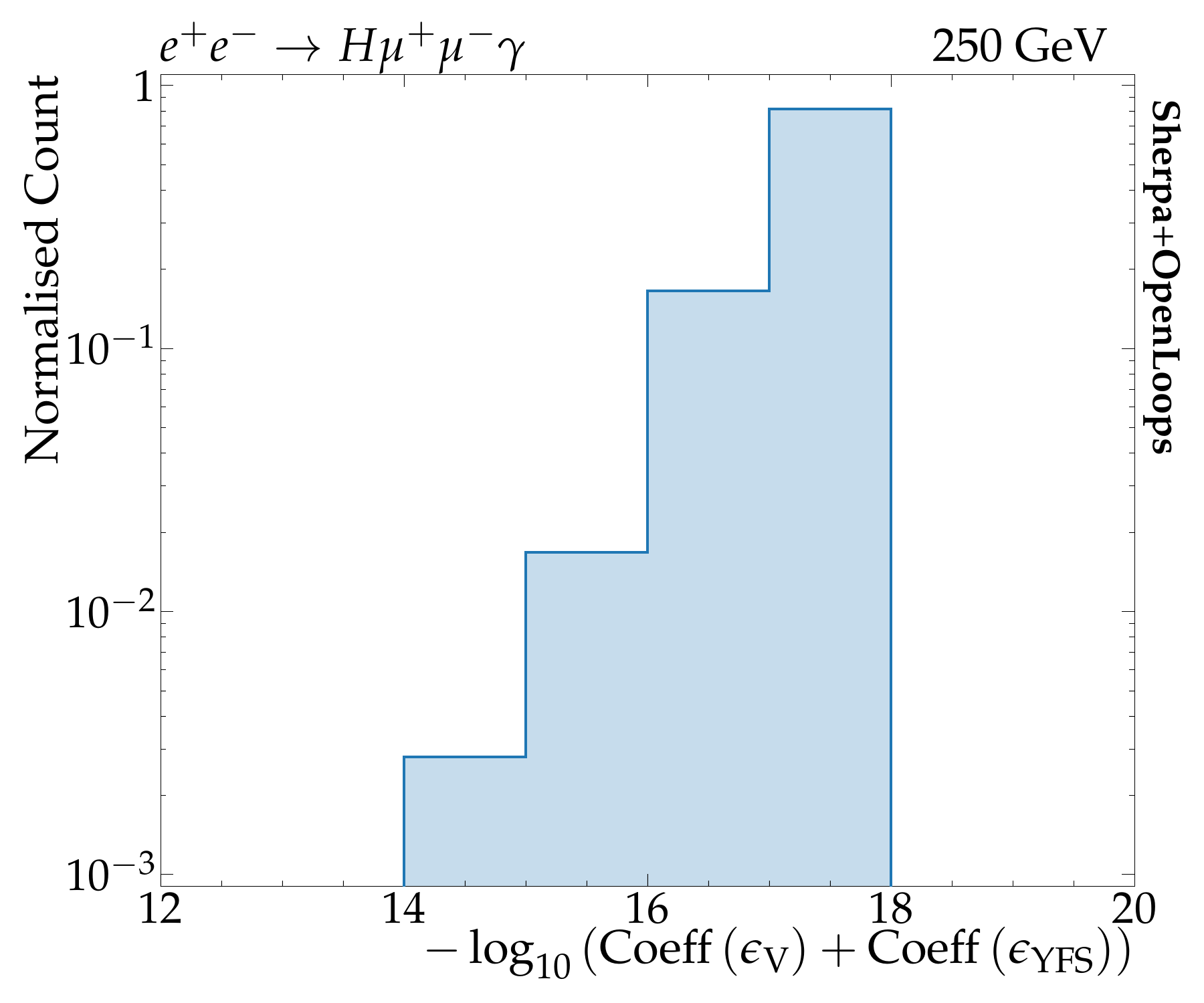}
    \includegraphics[width=0.4\textwidth]{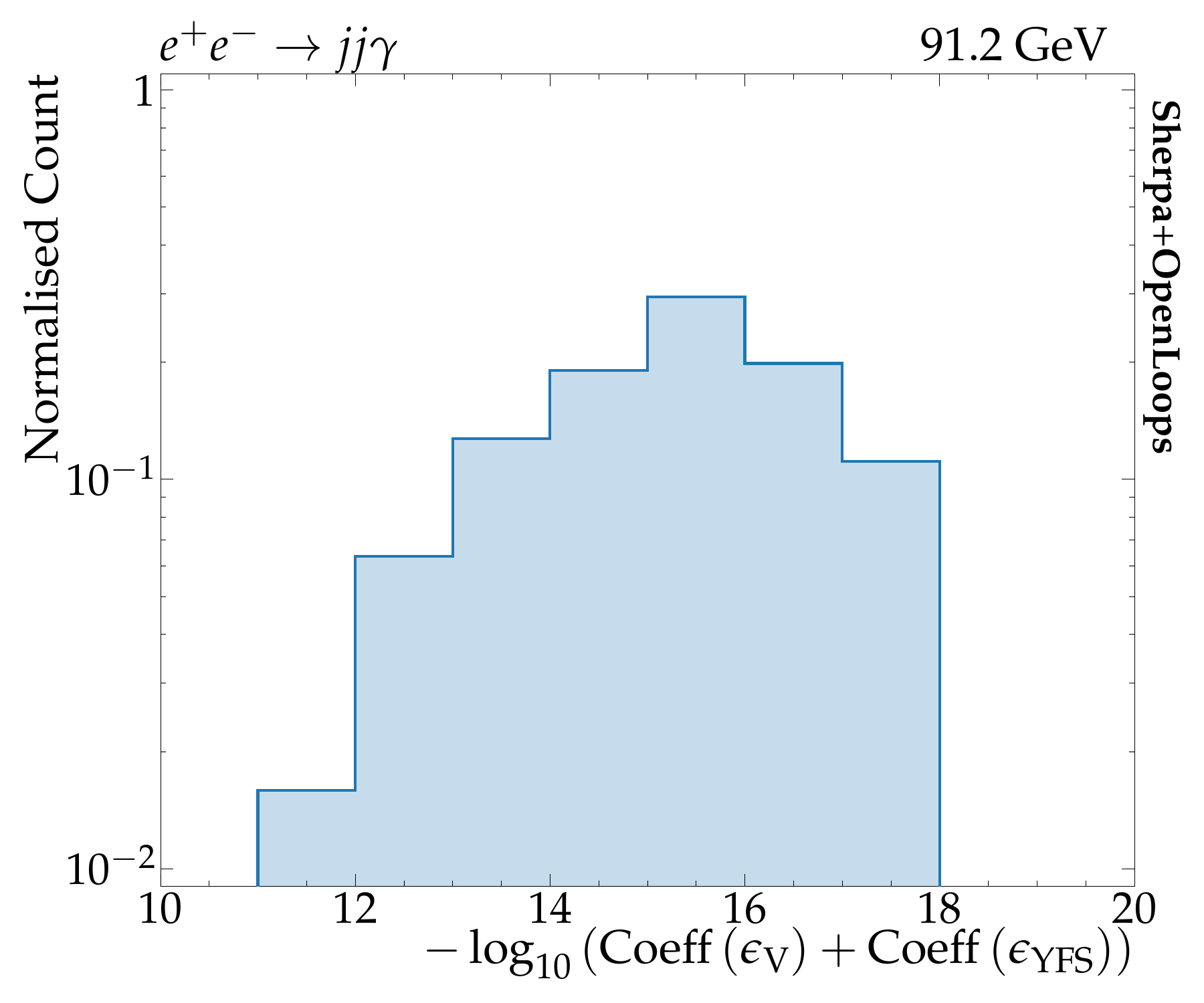}
    \includegraphics[width=0.4\textwidth]{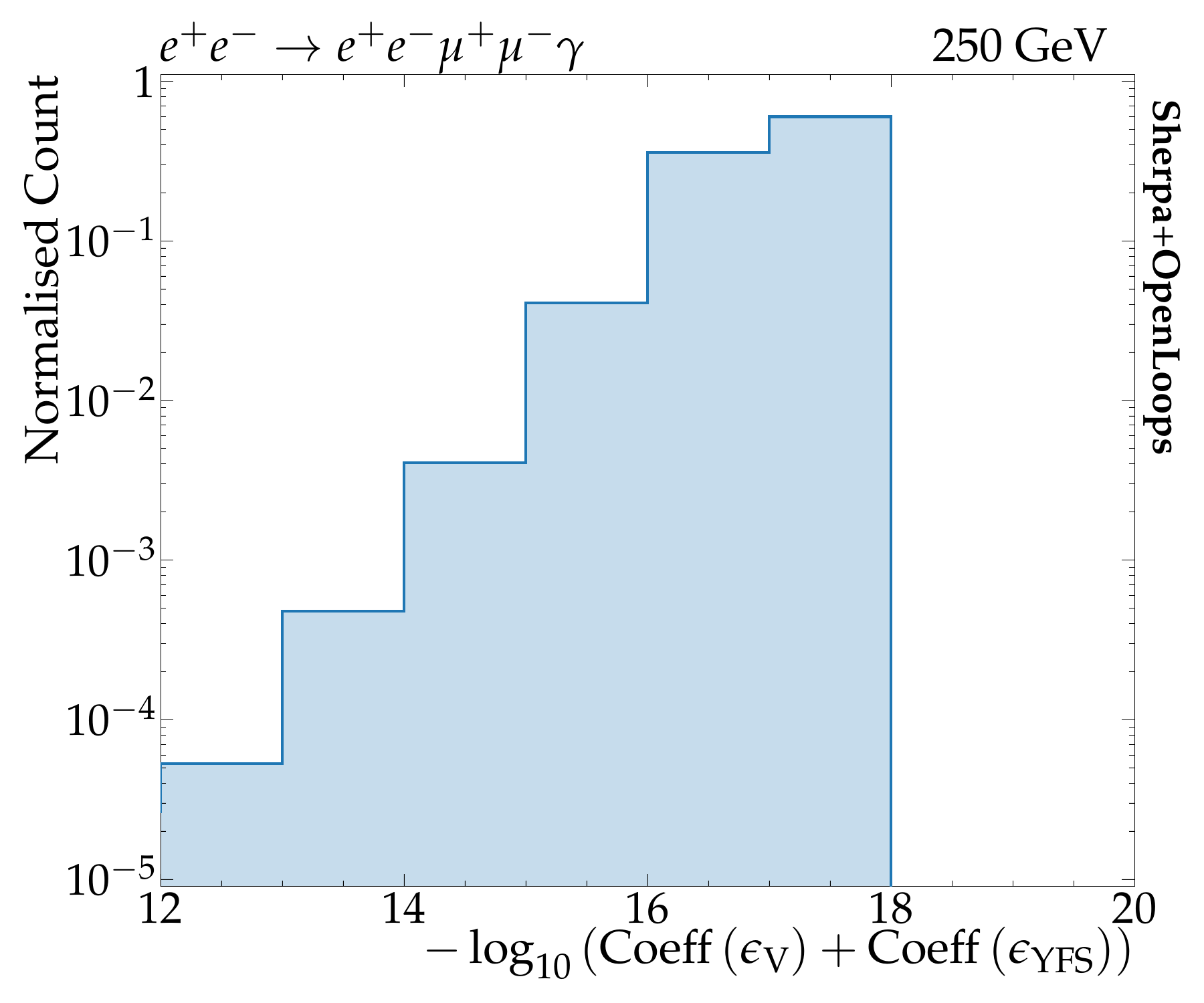}
    \caption{Cancellation of the IR poles according to~\cref{EQ:RealLoopIR} using dimensional regularization. 
    The bins correspond to the number of decimal places at which the cancellation was achieved.}
    \label{fig:RVSub}
\end{figure}

\noindent
In~\cref{fig:RVSub}, we demonstrate how the IR poles cancel according to~\cref{EQ:RealLoopIR} for our example processes. 
Once again, we observe that the IR poles agree to at least 14 digits, while in most cases, they cancel to 16 digits or better. 
Unsurprisingly due to the increased complexity, these results are slightly less stable than the ones in the one-loop case, but the cancellation is still well within the precision we can expect from double-precision machinery.
Although the $\mathcal{RV}(\Phi_{n+1})$ term is essentially a one-loop contribution, the presence of the additional real photon emission renders its numerical evaluation more challenging. 
For example, in the case of a four-particle born final state configuration, the real-virtual corrections consist of loop diagrams with seven external partons. 
Especially when including masses throughout the potentially large scale ratios pose a noticeable numerical precision challenge.

\subsubsection{Double Real Correction}\label{SubSec:RealReal}
\begin{figure}
    \centering
    \includegraphics[width=0.4\textwidth]{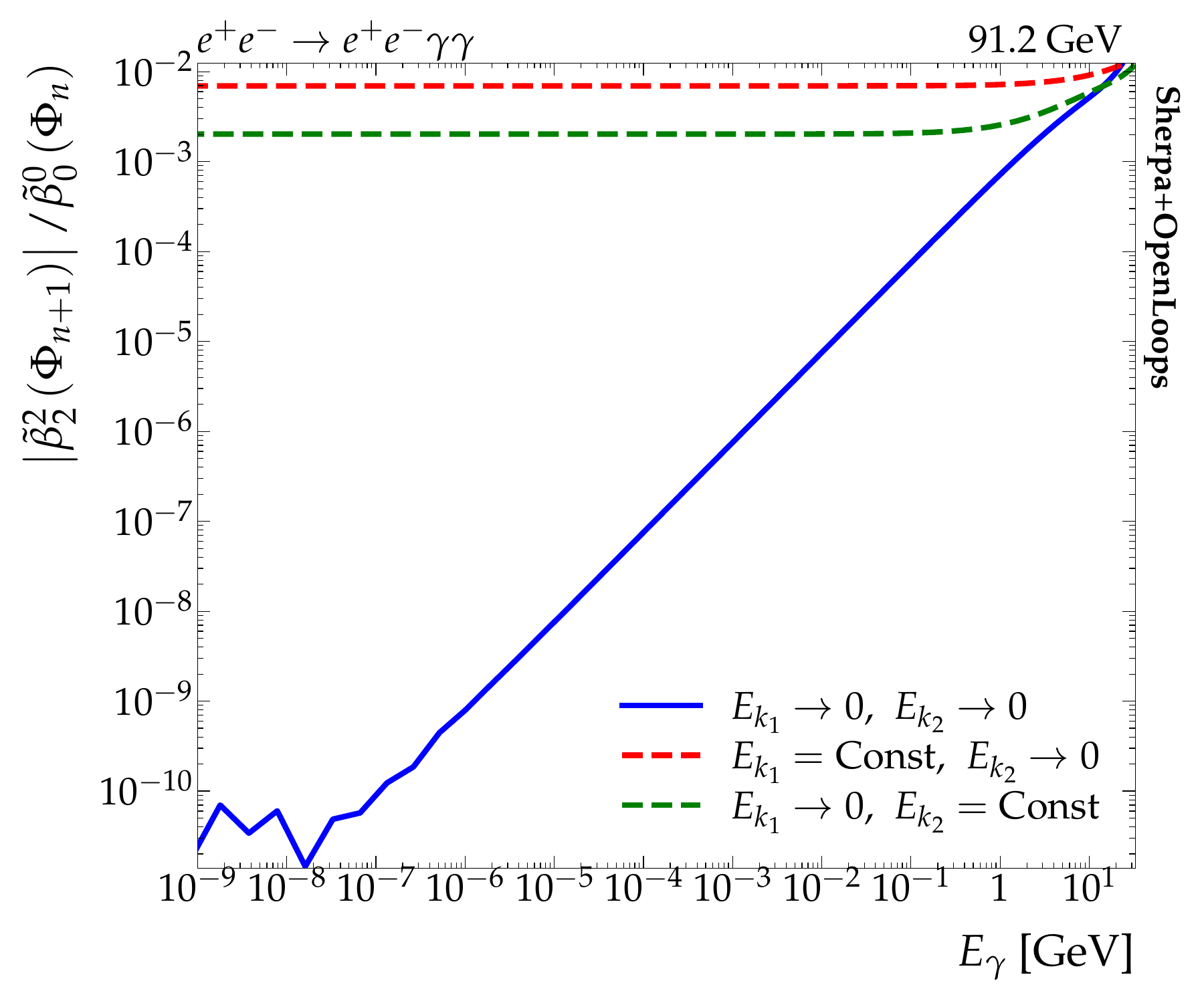}
    \includegraphics[width=0.4\textwidth]{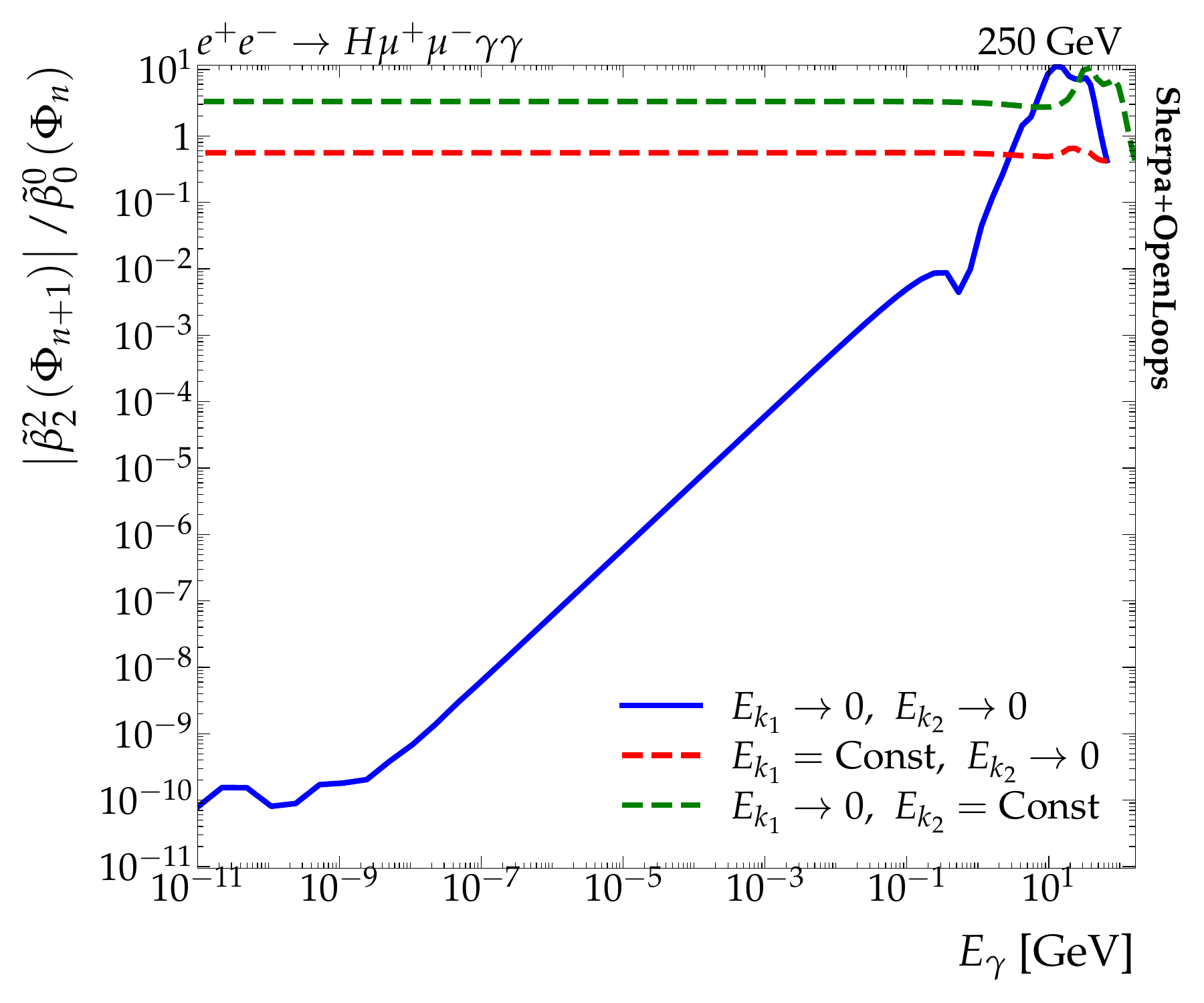}
    \includegraphics[width=0.4\textwidth]{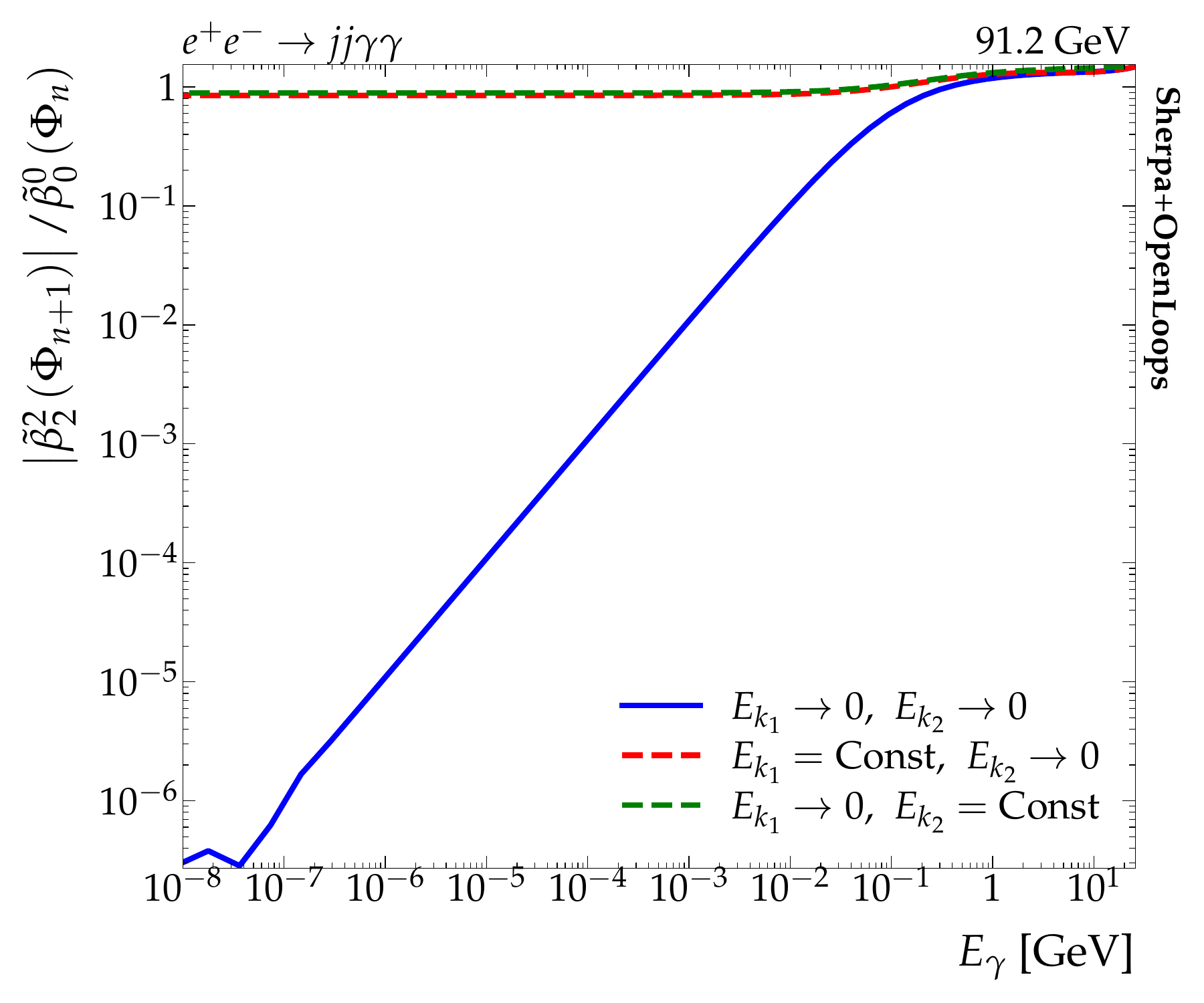}
     \includegraphics[width=0.4\textwidth]{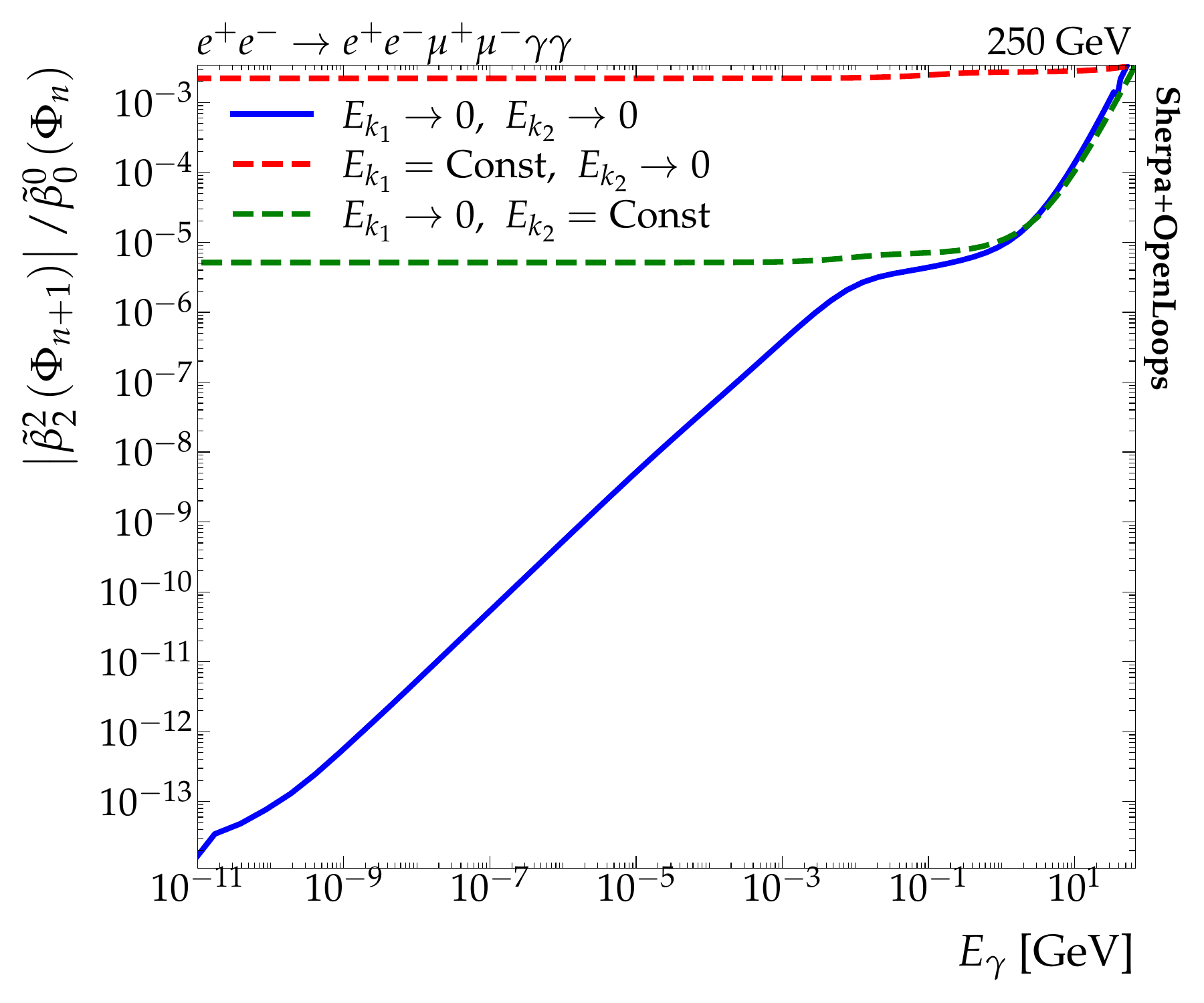}
    \caption{The scaling behaviour for our double real correction,~\cref{eq:RealRealSub}, normalized to the Born contribution, for the various processes.
    In the case where both photons become soft, we plot the results against the energy of the hardest photon on the x-axis. }
    \label{fig:RealRealSub}
\end{figure}

\noindent
The IR-finite contribution arising from the emission of two resolved photons can be written as,
\begin{align}\label{eq:RealRealSub}
\tilde{\beta}_2^2\left(\Phi_{n+2}\right) &= \mathcal{RR}\left(\Phi_{n+2}\right)
-\eik{1}\betatilde{1}{1}\left(\Phi_{n+1};k_2\right)
-\eik{2}\betatilde{1}{1}\left(\Phi_{n+1};k_1\right)
-\eik{1}\eik{2}\betatilde{0}{0}\left(\Phi_{n}\right),\\
\mathcal{RR}\left(\Phi_{n+2}\right) &=  \left(\frac{1}{2(2\pi)^3}\right)^2\left|\mathcal{M}^{1}_{2}\left(\Phi_{n+2}\right)\right|^2\,.
\end{align}
$\left|\mathcal{M}^{1}_{2}\left(\Phi_{n+2}\right)\right|^2$ marks the complete $n+2$ tree-level correction to the underlying born process, which can be calculated with standard automated tools. 
The remaining part of~\cref{eq:RealRealSub} represents the subtraction term for two photons, $k_1$ and $k_2$, with the dependence on the photon momentum made explicit. 
The first two terms in the subtraction consist of the IR-finite contribution from one photon, evaluated using~\cref{eq:RealSub}, multiplied by a local eikonal factor of the other photon.
The final term is simply the product of both eikonals with the Born term.
~\Cref{eq:RealRealSub} will have three kinematic regions in which we can expect the emergence of IR divergences: two of them where one photon becomes soft while the other remains hard, and one where both photons become soft. 
In the first two cases, where either $k_1$ or $k_2$ becomes soft and the remaining photon is fixed, the corresponding subtraction term will remove the singular behaviour from $\mathcal{RR}\left(\Phi_{n+2}\right)$, while the remaining subtraction term for the hard photon will simply leave us with a constant term plus a non-singular term from the squared amplitude. 
In the limit where both photons become soft, following the arguments presented in~\cref{SubSubSec:Real}, the contribution from $\tilde{\beta}_2^2\left(\Phi_{n+2}\right)$ becomes vanishingly small. 
We can test the first two limits by taking a phasespace point in which both photons are non-soft, keeping one photon constant while lowering the energy of the other and then evaluating~\cref{eq:RealRealSub} for each of these new phasespace points. 
To test the limit where both photons become soft, we apply this reduction procedure to both photons simultaneously. 
We show the results of this in~\cref{fig:RealRealSub}. 
In the single-photon IR-subtraction regions, the  contribution quickly reaches a constant, which is dominated by the hard contribution.
For the case where both photons are taken to the soft limit, the contribution becomes vanishingly small, again reflecting the fact that the contributions in this region of phasespace have already been resummed. 

\noindent
We find that our implementation of YFS subtraction is extremely stable for ultra-soft double photon emissions across a wide indicative range of processes. 
It is also worth noting that the double-real emission represents the first non-trivial dipole structure for our subtraction as we have to consider photon emissions that can be both from the initial or final state as well as one emission from the initial and the other from the final state. 
The fact that our subtraction successfully removes all IR divergences for such emissions, for arbitrary processes, implies that our approach can be extended even beyond \NNLOEW accuracy as the subtraction terms for the extra resolved emissions will still only involve three types of dipoles, namely initial, final, and initial-final. 
Given that our momentum mappings are stable for each dipole and that the addition of more resolved emissions will not affect our algorithm, as we will show in~\Cref{SubSec:MMaps}, we expect that the tree-level corrections beyond \NNLOEW will not be a limiting factor for us.

\subsubsection{Double-Virtual Correction}\label{SubSec:VV}
\noindent
The final ingredient to complete our \NNLOEW corrections are the two-loop contributions, which are among the most complicated contributions that have to be calculated for the matching of the full \NNLOEW corrections. 
While there has been significant progress in the calculation of two-loop amplitudes for \ee~\citetwoloop processes, there is currently no tool publicly available for the automated calculation of two-loop amplitudes. 

The subtraction term for the two-loop contribution is given by,
\begin{equation}\label{EQ:TwoLoopIR}
    \betatilde{0}{2}\left(\Phi_n\right) = \mathcal{VV}(\Phi_n) - \sum_{ij}\mathcal{D}_{ij}\left(\Phi_{ij}\otimes\Phi_n\right)\betatilde{0}{1}\left(\Phi_n\right) - \frac{1}{2}\left(\sum_{ij}\mathcal{D}_{ij}\left(\Phi_{ij}\otimes\Phi_n\right)\right)^2\betatilde{0}{0}\left(\Phi_n\right).
\end{equation} 
where $\mathcal{VV}(\Phi_n)$ is the complete two-loop correction. 
The remaining terms represent the YFS subtraction term, which will contain both single and double poles in $\epsilon$. 
This will be the first appearance of double poles in our calculation, as we have regulated the collinear divergences with our non-zero fermion masses such terms have not appeared before now.
The second term is similar to the one-loop subtraction term, however the Born contribution \betatilde{0}{0} has been promoted to the IR-finite one-loop correction~\cref{EQ:OneLoopIR}. 
The final term is the squared subtraction term, where the double poles will appear as well as additional singular contributions. 

\begin{figure}
    \centering
    \includegraphics[width=0.45\textwidth]{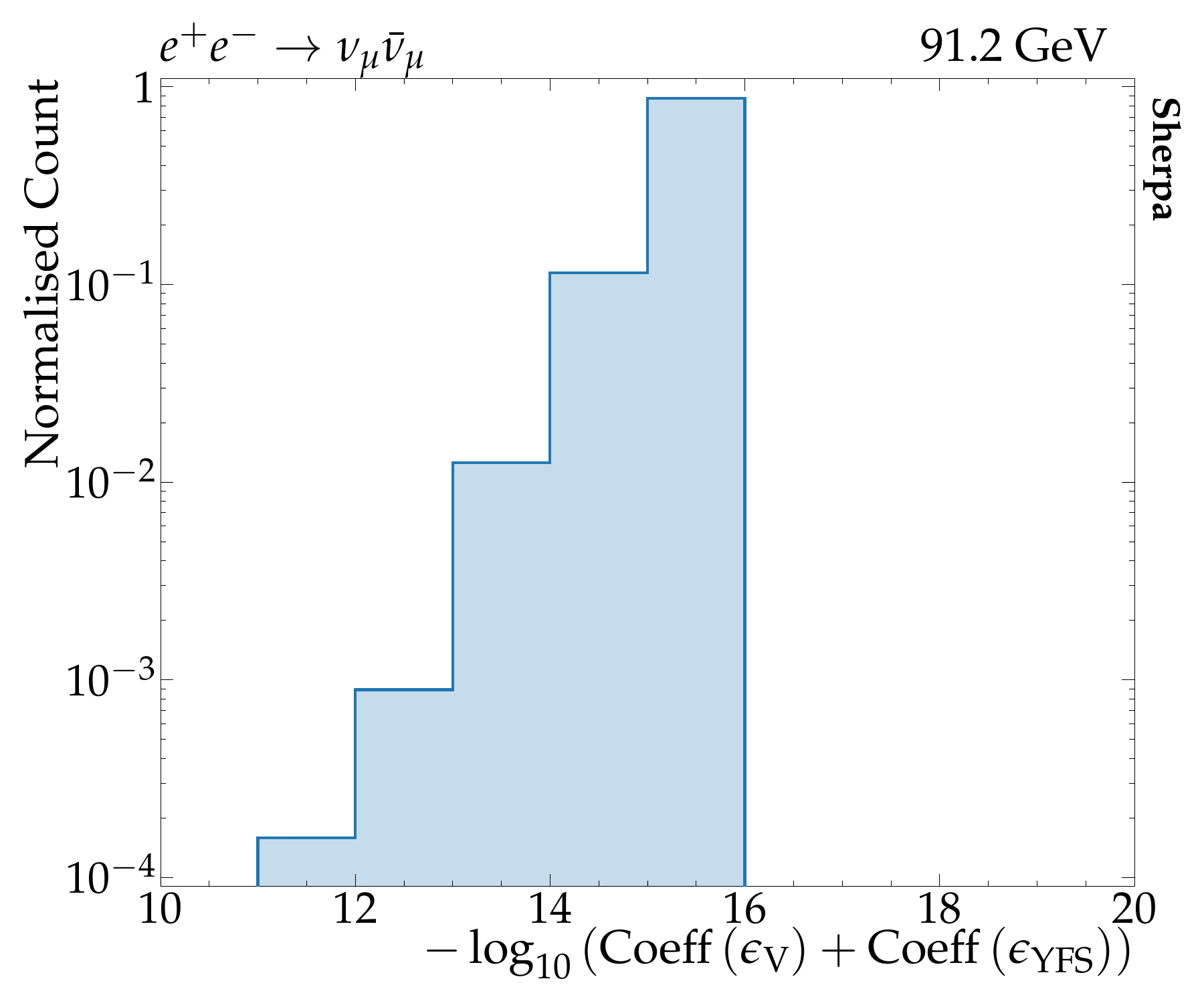}
    \includegraphics[width=0.45\textwidth]{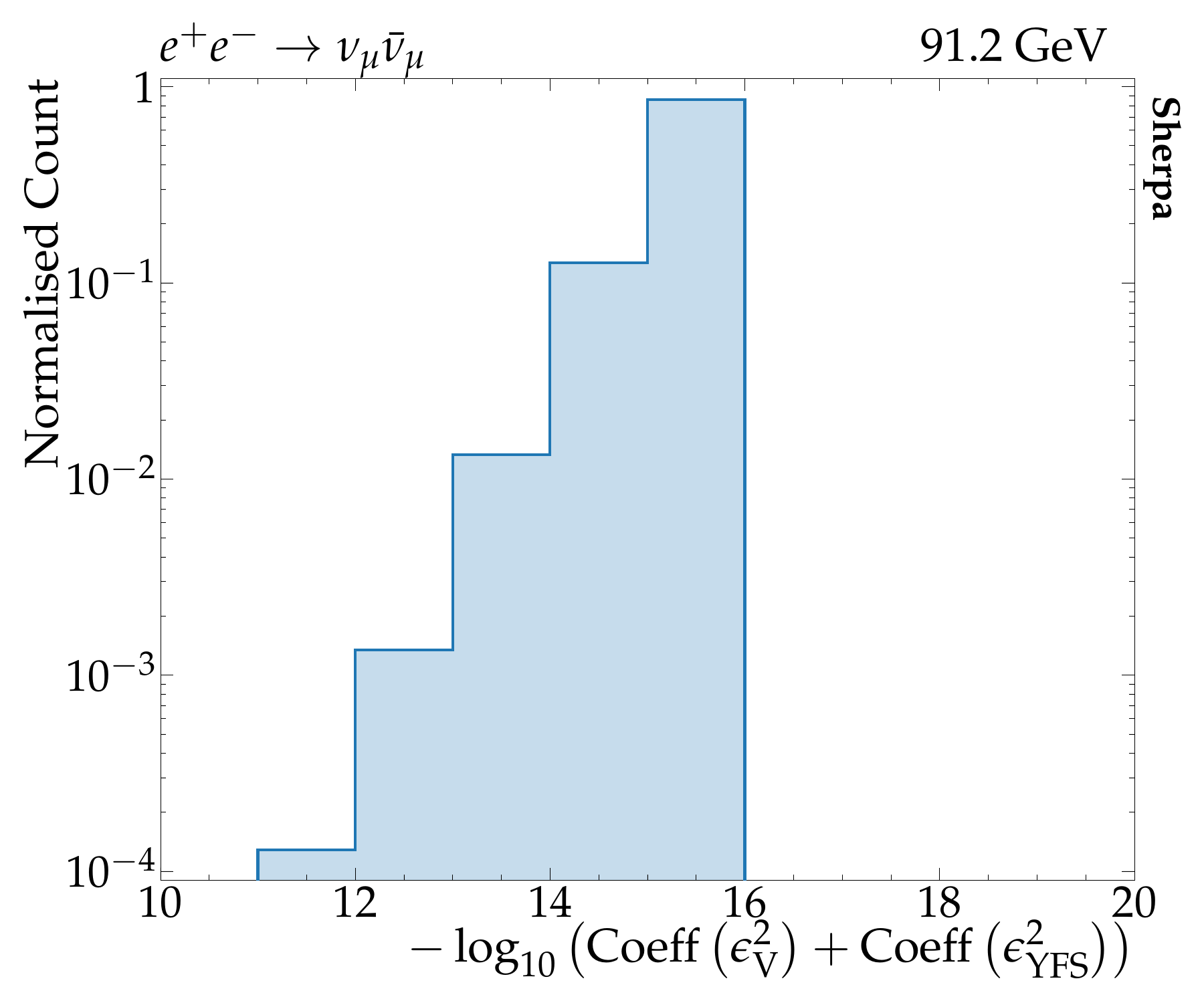}
    \caption{Cancellation of the IR poles according to ~\cref{EQ:TwoLoopIR} using dimensional regularization for both the single(left) and double(right) poles. 
    The bins correspond to the number of decimal places at which the cancellation was achieved.}
    \label{fig:VVSub}
\end{figure}

\noindent
The subtraction works for the final state, as already discussed in~\cite{Krauss:2018djz}; we checked our reimplementation by considering the two-loop QED corrections to $\nu_{\mu} \bar{\nu}_{\mu} \rightarrow \ee$. 
To test our implementation of YFS subtraction for initial state dipoles we have implemented the two-loop  initial state radiation (ISR) QED corrections~\cite{Berends:1987ab,Blumlein:2020jrf} to neutrino production in dimensional regularization; in~\cref{fig:VVSub} we show both the single and double pole cancellation. 
Remaining corrections, in particular the IR-finite weak corrections, can be easily included as and when they become available in a format suitable for YFS.

\noindent
One set of two-loop amplitudes that we have at our disposal are those that are provided by the \Griffin package~\cite{Chen:2022dow}, which we have interfaced directly to \Sherpa.
While not only providing our missing \NNLOEW contributions at the Z-Pole, \Griffin also includes additional higher-order corrections, including the complete \oforder{\alpha\alpha_s} and \oforder{\alpha\alpha_s^2} corrections to the EW form factors as well as the leading higher-order contributions extracted from an expansion in the top mass.
\Griffin also removes the IR singularities with a YFS inspired subtraction, which means that in our interface we do not have to supply the corresponding subtraction terms.
As a consequence, we carefully checked our interface to ensure that we correctly account for all IR-finite contributions, resolving subtle mismatches between our respective subtraction schemes.
This was achieved by comparing the \Griffin results to our IR-subtracted results from tools such as \OpenLoops and \Recola, with the remaining IR-finite QED contributions that have been previously implemented in \Sherpa~\cite{Krauss:2022ajk,Jadach:2000ir}.

\noindent

\subsection{Momentum Mappings}\label{SubSec:MMaps}
\noindent
In this section, we will detail the momentum mappings that are crucial in constructing our higher-order corrections. 
It is important to note that the momentum mappings are used to construct higher-order corrections only and that they will only be included in the event record in cases where both the local and global phasespaces coincide. 
By global we refer to the full event phasespace, i.e. the set of momenta included in the recorded event, including all photons. By local, we refer to the reduced set of momenta evaluated at the boundary of the global phasespace, where one or more photon momenta vanish, as required when evaluating a real residual 
\betatilde{}{}. This local phasespace is introduced to facilitate the evaluation of real corrections and does not correspond to any physical event record.
For example, in cases where there is exactly one photon emission, the real-correction phasespace coincides with the full event phasespace. In essence, we reapply the traditional YFS phase-space mappings, but restrict them to include only the momentum of the photon to which we match the real-emission corrections

\noindent
First, we need a mapping that takes the complete phasespace $\Phi_{n+n_k}$ and maps it to phasespaces of equal or less dimension $\Phi_{n+n_R}$. 
As before, $n_R$ is the number of resolved photons we are correcting for. 
The second set of mappings projects the dipole phasespace, $\Phi_{\tilde{i}\tilde{j}}$, onto the phasespace where all photon momenta vanish except for the one we are matching to.
Crucially, the dipoles appearing in the subtraction terms are mapped from the global phasespace to the corresponding local one, where the subtraction occurs.
Since QED is silent on the ordering of photon emissions for a given dipole, we treat the emissions in a completely democratic fashion: we do not identify individual emitters and spectators, but instead allow the combination of all particles to absorb the recoil of the emission. 

\noindent
Let us first consider the emission of an arbitrary number of photons from the initial state dipole. 
We first construct a boost frame, $\mathcal{P}$,
\begin{equation}
    \mathcal{P} = p_i+p_j+K\,.
\end{equation}
Here, the beam momenta $p_{i,j}$ are the momenta of the initial dipole after ISR defined through,
\begin{equation}
    \tilde{p_i}+\tilde{p_j} \rightarrow  p_i + p_j + K,
\end{equation}
and $K$ is the sum of matched photon momenta.
In the case of~\cref{eq:RealSub}, $K=k_1$, while for ~\cref{eq:RealRealSub} $K=k_1+k_2$ and so on. 
While it is not required that $K=\sum_{l} k_l$, this can arise, for example, when constructing the $\betatilde{1}{1}$ correction to an event with only one resolved photon emission.
We then boost and rotate the four-momenta of the final state particles to this frame and introduce a parametrization of the initial state dipole momentum as,
\begin{equation}{\label{Eq:InitialMap}}
    p_{i,j} = (E_{i,j}, 0, 0, \pm p^Z), \,\, E_{i,j} = \sqrt{m_{i,j}^2 + (p^Z)^2 }, \,\, p^Z =\frac{1}{2}\sqrt{\frac{\lambda(s_P,m_i^2,m_j^2)}{s_Q}},
\end{equation}
where $\lambda$ is the K\"all\'en function and 
\begin{equation}
    s_Q = (p_i+p_j+K)^2.
\end{equation}
After this, we boost all final state momenta back to the lab frame defined by $s_Q$. 
This final transformation will leave us with the phasespace point that will enter the amplitude calculation. 
Since we will only ever have one dipole from the initial state the mapping for the dipole term, in particular the dipole entering the eikonal factors, will be the $p_{i/j}$ as defined above. 
This method can be applied to any number of initial state photons, and the inclusion of virtual emissions, as in~\cref{EQ:RVSub}, does not require us to adapt the mappings.

\noindent
For emissions from final state dipoles, the situation is more complicated because there could be more than one unique dipole. 
While we still need to ensure we are providing the amplitude with the correctly mapped phasespace point, we must consistently transform each dipole to guarantee that we remove the IR singularities. 
The first step is to construct all possible final state dipoles from charged particles only. 
The initial momenta in these dipoles are taken to be the Born-level momenta, i.e.\ the momenta before any FSR emissions. 
Thus, the dipole momenta are initially defined through $\tilde{q}_{i,j}$ as,
\begin{equation}
     \tilde{q_i}+\tilde{q_j} \textbf{}\rightarrow  q_i+q_j + \sum_{m=1}^{N_{\text{FSR}}} k_m.
\end{equation}
We first boost the momenta to the rest frame of the dipole, 
\begin{equation}
    \mathcal{Q} = (s^\prime_Q, 0, 0, 0), \,\, s^\prime_Q = \frac{s_q}{1+\frac{2K^0}{\sqrt{s_q}}}, \,\, s_q = (\tilde{q_i}+\tilde{q_j})^2\,.
\end{equation}
In this frame, the dipole momenta are,
\begin{equation}
    q_{i,j} = (E_{i,j}, 0, 0, \pm p^Z), \,\, E_{i,j} = \sqrt{m_{i,j}^2 + \left(p^Z\right)^2 }, \,\, p^Z =\frac{1}{2}\sqrt{\frac{\lambda(s^\prime_Q,m_i^2,m_j^2)}{s_Q}}.
\end{equation}
Next, we boost the momenta to a new "after-emission" frame, $\mathcal{Q}_K=\mathcal{Q}+K$. 
Then we boost the new dipole momenta and photons back to the original lab frame. 
This mapping is performed separately for each dipole entering the subtraction, such that each eikonal term is constructed from local dipole momenta, which will differ from the global momenta which enter in explicit eikonals in~\cref{eq:masterYFS}. 
Once all the final state dipoles are in this frame we perform one last mapping on the initial state dipole, identical to~\cref{Eq:InitialMap}, where $S_Q$ is taken as the sum of all final state momenta.

For any number of initial or final state photon emissions the mappings can proceed as described above. 
This leaves us with the most intricate case where we have both an initial and final state particle emitting photons.
In such cases we first apply the final state mappings, then by incorporating the final state momenta, including all final state photons, we perform the initial state mappings. 
One final phasespace correction needs to be constructed, which in essence is a pseduo-flux factor~\cite{Jadach:2000ir}.
Depending on the dipole type, we correct them with,
\begin{equation}\label{EQ:Flux}
    \mathcal{F}\left(k_1,\cdots,k_n\right) = \frac{\left(Q-k_1+\cdots+k_n\right)^2}{Q^2}
\end{equation}
where $Q$ is the sum of the dipole momenta, and we take into account only the $n$ real photons that we are matching. 
Note that such a kinematical factor is already present in both the fixed-order and leading-log approximations, as well as in the underlying crude YFS distribution~\cite{Jadach:1988gb}.
The inclusion of this factor in the real corrections, such as~\cref{eq:RealSub,eq:RealRealSub}, order by order improves the overall stability of our calculations, while leaving the resummation untouched. 
As in the soft limit of vanishing photon momenta~\cref{EQ::FormFactor} is simply unity and has no effect.

\section{Results}

\begin{table}
\setlength{\tabcolsep}{11pt}
\renewcommand{\arraystretch}{1.5}
  \begin{center}
    \begin{tabular}{ c| c | c}
    LO [fb] & \yfsnlo [fb] & $\mathrm{PDF}_\mathrm{LL} + \mathrm{NLO}_{\mathrm{EW}}$ [fb]\\
      \hline
      \hline
      1.593(1) & 1.361(5) & 1.368(9) \\
     \hline
    \end{tabular} 
    \parbox{0.8\textwidth}{\caption{\label{tab:fid-xs}Fiducial cross-sections for
     $\ee \rightarrow \mu^+\mu^-\tau^+\tau^-$ at 250~\UGeV. The Monte-Carlo uncertainties are quoted within the brackets.}}
  \end{center}
\end{table}

\begin{figure}[h]
    \centering
    \includegraphics[width=0.45\linewidth]{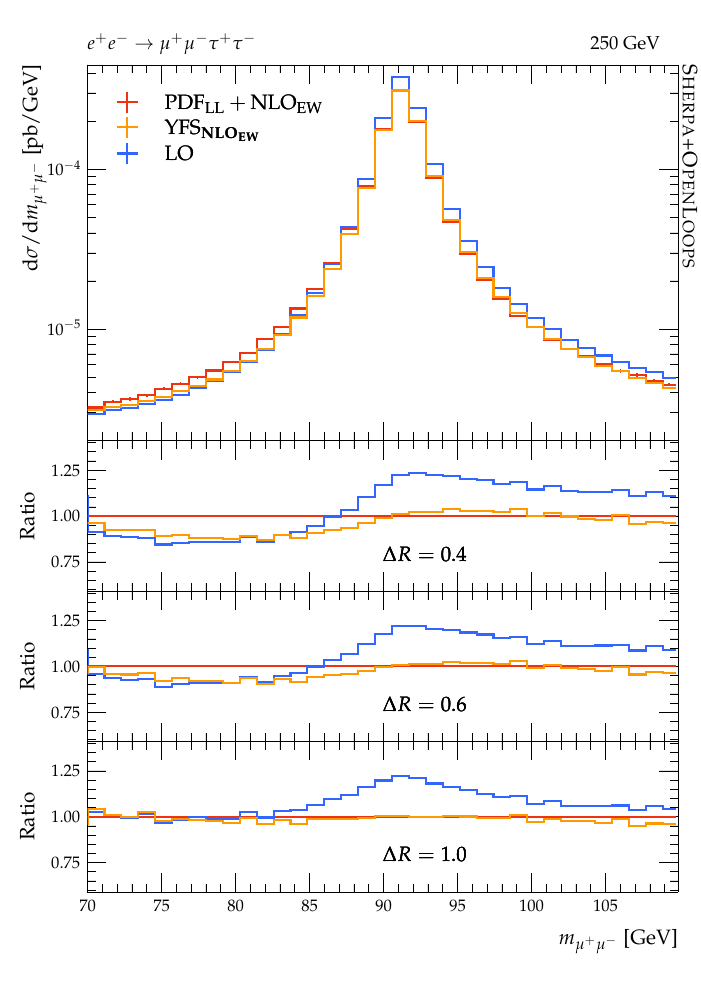}
    \includegraphics[width=0.45\linewidth]{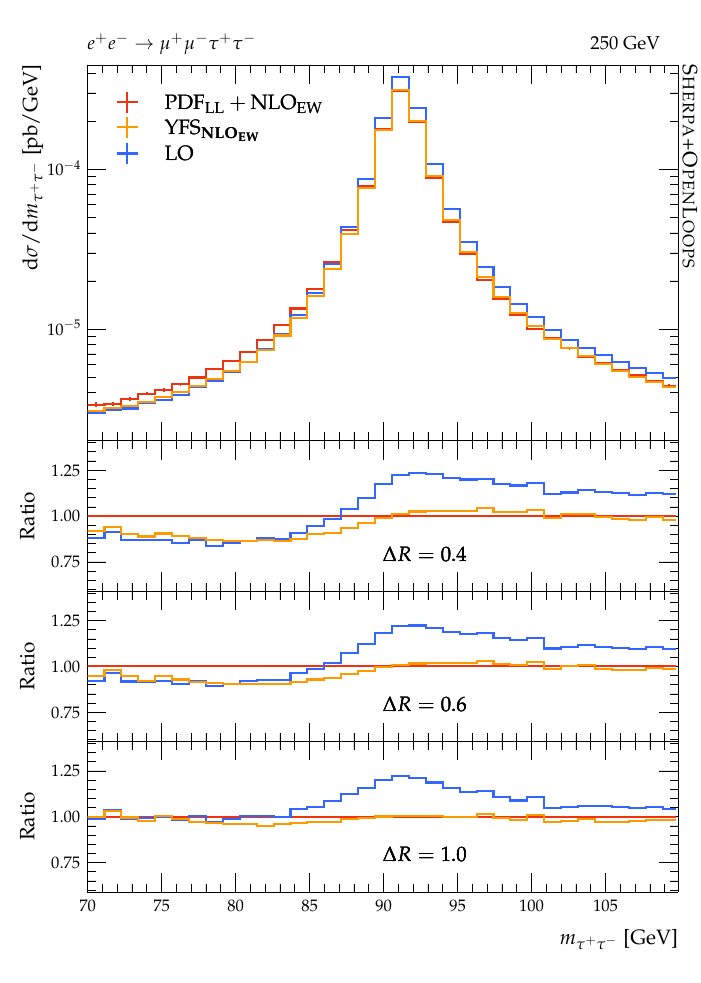}
    \caption{ Invariant dilepton mass distributions for same flavour final states. 
    The nominal plots are taken with $\Delta R = 0.4$}
    \label{fig:ZMass}
\end{figure}

\begin{figure*}
    \centering
     \includegraphics[width=8.5cm]{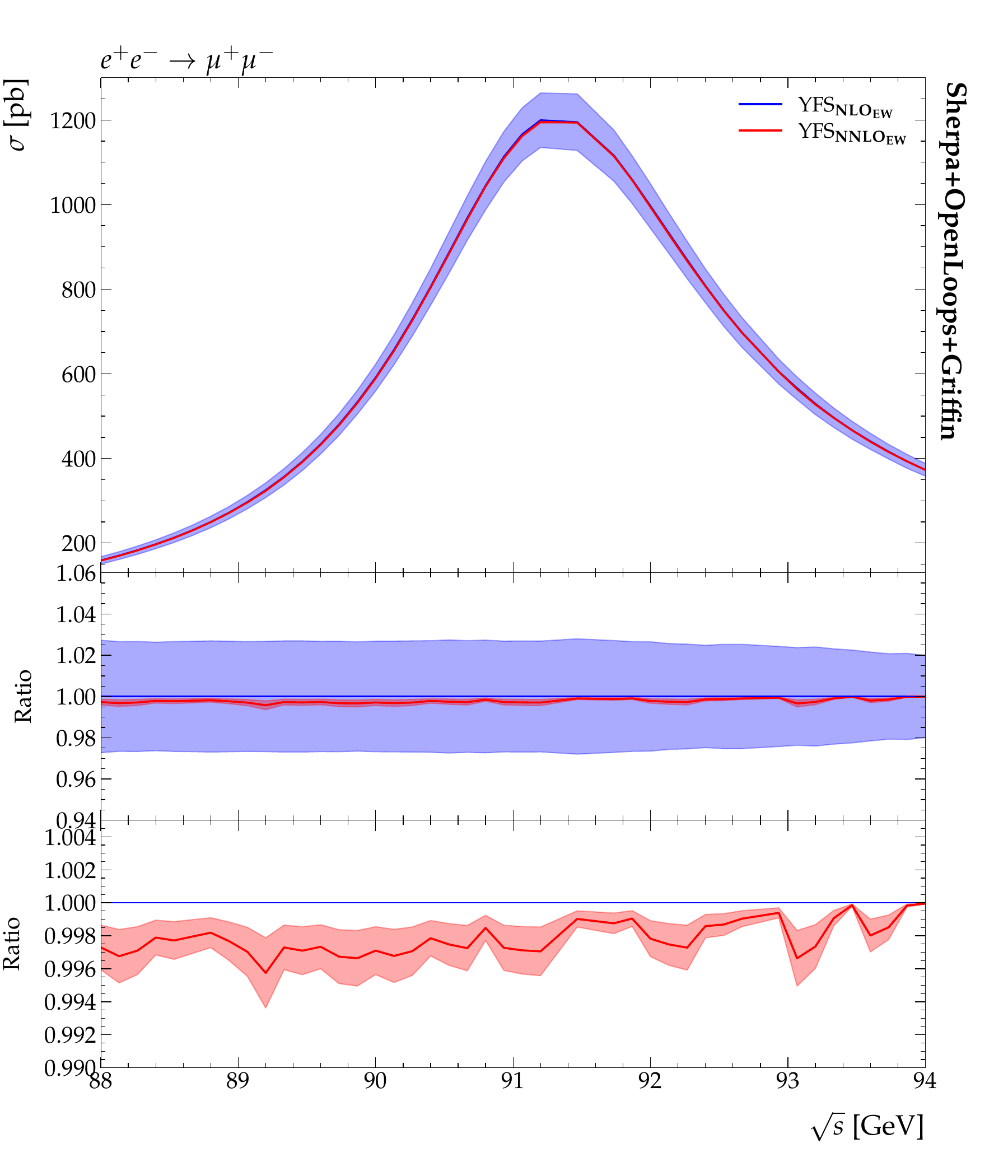}
    \caption{Total calculated cross-section for muon pair production at energies around the Z-pole. 
    Both the \yfsnlo (blue) and the \yfsnnlo (red) results are presented along with their associated perturbative uncertainties. }
    \label{fig:TotalXS}
\end{figure*}

\begin{figure}
    \centering
     \includegraphics[width=0.4\textwidth]{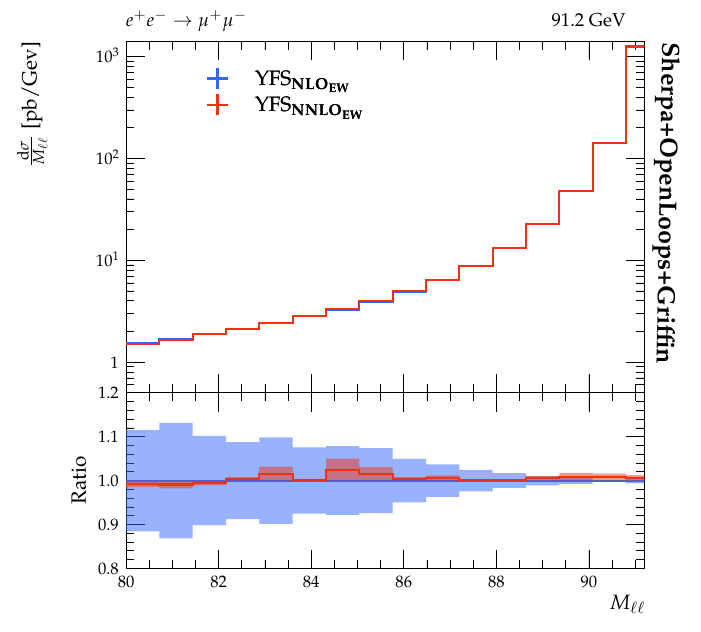}
     \includegraphics[width=0.4\textwidth]{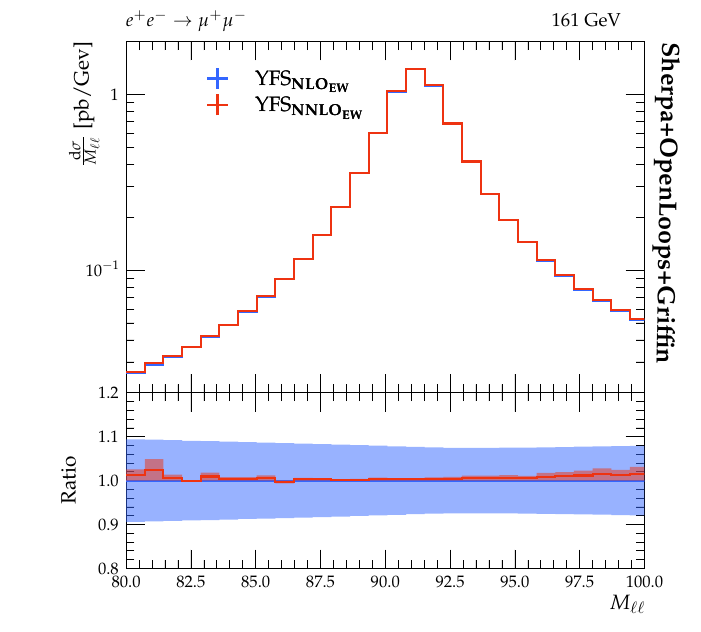}
     \includegraphics[width=0.4\textwidth]{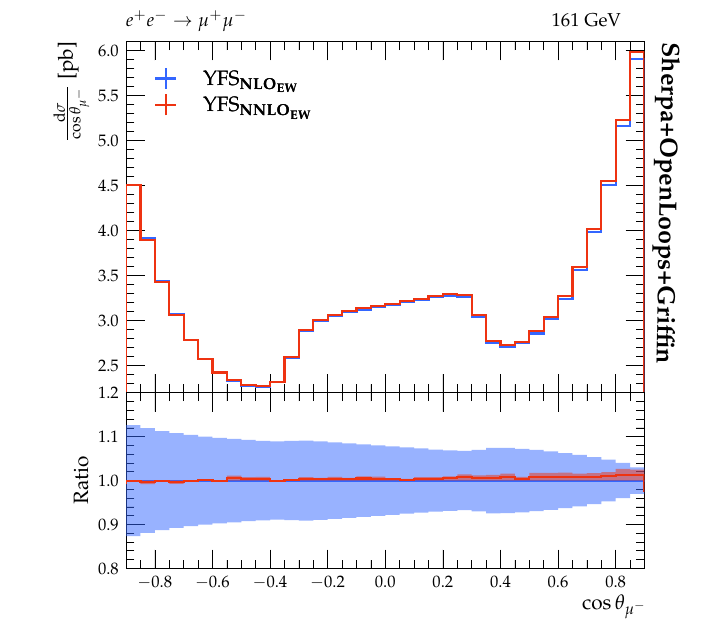}
     \includegraphics[width=0.4\textwidth]{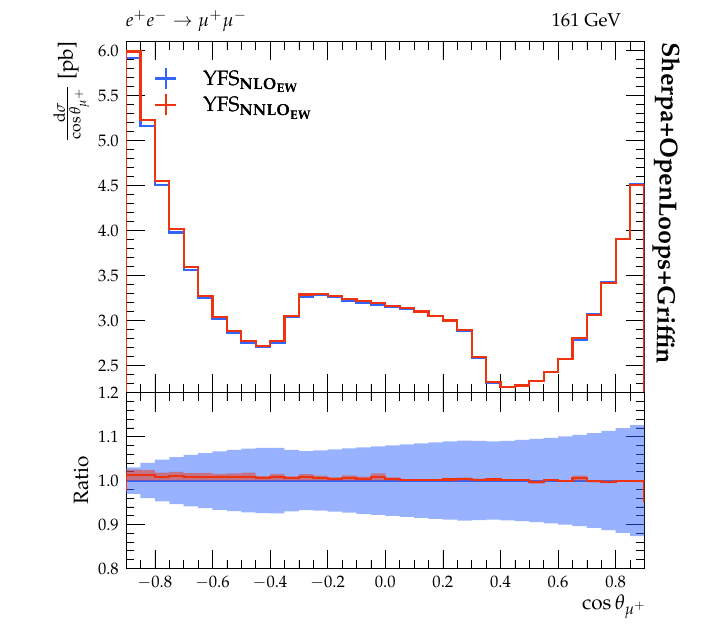}
    \caption{Differential muon pair invariant mass distributions at 91.2~\UGeV(left) and 161~\UGeV(right). 
    Both the \yfsnlo (blue) and the \yfsnnlo (red) are presented along with their associated perturbative uncertainties. }
    \label{fig:InvariantMass}
\end{figure}

\begin{table}
\setlength{\tabcolsep}{11pt}
\renewcommand{\arraystretch}{1.5}
  \begin{center}
    \begin{tabular}{c| c | c}
      & Mass [ \UGeV] & Width [ \UGeV] \\
      \hline
      \hline
         Z & 91.1876 & 2.4952 \\
         W & 80.379 & 2.085\\
         e & 0.000511  & - \\
     $\mu$ & 0.105 & - \\
     $\tau$ & 1.777 & -\\
     \hline
     \hline
     $\alpha^{-1}\left( 0 \right)$ & 137.03599976
    \end{tabular}
    \parbox{0.8\textwidth}{\caption{\label{tab:Validation:ew-inputs} 
        Electroweak input parameters.}}
  \end{center}
\end{table}

\noindent
Before we consider the new results of our implementation, we shall first consider
a validation setup, where we compare our approach with \Sherpa's fixed-order predictions,
utilizing our automated Catani-Seymour (CS) subtraction~\cite{Schonherr:2017qcj}. We will
consider the four-fermion process $\ee \rightarrow \mu^+\mu^-\tau^+\tau^-$ at 250~\UGeV.
We apply the following fiducial phasespace cuts,

\begin{equation}
p_T^\ell \geq 20\,\UGeV \,\,\quad 70\,\UGeV \leq M_{\ell\bar{\ell}} \leq 110\, \UGeV.
\end{equation}
In addition, we also require at least one isolated photon~\cite{Frixione:1998jh} with 
$p_T^\gamma \geq 1\,\UGeV$. We also apply an electron PDF in the leading log 
approximation for the CS setup. We employ the $G_\mu$ input scheme where
the EW couplings are derived from the Fermi-constant and gauge-boson masses via,
\beq
\alpha = \left|\frac{\sqrt{2}\sinSqW\mu_W^2 G_\mu}{\pi}\right|,
\eeq
where $\mu_i$ are the complex-valued renomalized masses~\cite{Denner:2005fg}
which are also used to derived the weak mixing angle,
\beq
\mu_i^2 = M_i^2 - i\Gamma_iM_i,\,\quad \sinSqW = 1-\frac{\mu_W^2}{\mu_Z^2}
\eeq
where the numerical values of the masses are given in~\cref{tab:Validation:ew-inputs}.
Since there is currently no publicly available matched generator that can combine 
a QED shower with a fully differential \NLOEW calculation, we apply a variety 
of different recombination radii to the final-state charged leptons, clustering 
collinear photon emissions that lie within a cone of radius $R$ around the lepton. 
The fiducial cross-sections are given in~\cref{tab:fid-xs}.
We observe excellent agreement between the \yfsnlo and the CS approaches. 
The \NLOEW corrections induce an approximately $14\%$ reduction in the fiducial
cross-section compared to the leading-order predictions. In~\cref{fig:ZMass}, we
show the invariant mass distributions for both same-flavour final states, 
evaluated using three different choices of recombination radius $\Delta R$. We 
find that as $\Delta R$ is increased to unity, the agreement between the two 
approaches improves, particularly below the Z-pole. In this limit, the effect 
of multi-photon radiation, present in \yfsnlo but not in $\mathrm{PDF}\mathrm{LL}
+ \mathrm{NLO}{\mathrm{EW}}$, is reduced, such that the two predictions coincide.
We also note a slightly larger deviation for the final-state $\tau$'s, which we 
attribute to mass effects, as our CS subtraction is currently limited to massless leptons in \ee setups.

\noindent
We turn to present results up to \NNLOEW accuracy for $e^+e^- \rightarrow f \bar{f}$ where $f\neq e$. 
As a first example, we consider the production of a muon pair. 
In~\cref{fig:TotalXS}, we show the total cross-section for this process over an energy scan of the Z-pole, both for \yfsnlo and \yfsnnlo. 
The former only includes corrections \oforder{\aew},  while the latter also includes the complete \oforder{\aew^2} contributions. 
The uncertainty bands refer to our estimate of missing higher-order corrections and are constructed by considering the difference of the nominal value with the next lowest order prediction along the procedure outlined in~\cite{Jadach:2000ir,Grunewald:2000ju} and our uncertainty envelopes are given by,
\begin{equation}\label{EQ:Uncert}
    \Delta \mathrm{YFS}_{i} = \left(\frac{\mathrm{YFS}_{i}-\mathrm{YFS}_{i-1}}{\mathrm{YFS}_{i-1}}\right) \quad\quad  i\in (\LO,\NLOEW,\NNLOEW).
\end{equation}
We also note that this is the first time the complete \yfsnnlo has been calculated for this process: we have included the missing contributions, notably the IR finite contributions, arising from the real-virtual corrections, which so far have been absent from the KKMC program~\cite{Jadach:2022mbe}. 
We see that the uncertainty of the \yfsnlo corrections remains quite stable at $\approx2.5\%$ while for the \yfsnnlo we see a reduction of the uncertainty to levels of $\approx0.1\%$.
The large size of the \yfsnlo error band reflects the fact that the \yfslo contains no higher-order corrections; in particular it lacks any corrections due to collinear logarithms.
This reduction highlights the stability of our perturbative expansion, while the full theoretical uncertainty will require further exploration of parametric effects.
To assess them we need to compute the amplitudes with different electroweak parameters.
\noindent
In~\cref{fig:InvariantMass} we show the differential distribution for the invariant mass of the final state at two different energies, the $Z$-pole and $WW$ threshold. 
In the first case, we see that near the Z-pole the uncertainty associated with \yfsnlo is negligible, reflecting the fact that most of the effects have been accounted for in the resummation. In contrast, as we move away from the $Z$-pole to regions where we have a hard photon emissions the uncertainty estimate begins to increase, reflecting the fact that the \yfslo predictions fail to correct for such emissions. 
Going to \yfsnnlo we see the difference drop satisfyingly to the per-mil level when compared to \yfsnlo as reflected in the uncertainty bars. 
At the $W$W threshold, we see considerably larger uncertainties for the \yfsnlo case; again, this reflects that we have a large radiative return to the $Z$-pole that must be accounted for. 
Here the \yfsnnlo is in good agreement while above the $Z$-pole the error increases somewhat, indicating potentially large initial-final interference effects in the \oforder{\aew^2} corrections.


\begin{figure}[h]
    \centering
    \includegraphics[width=0.42\linewidth]{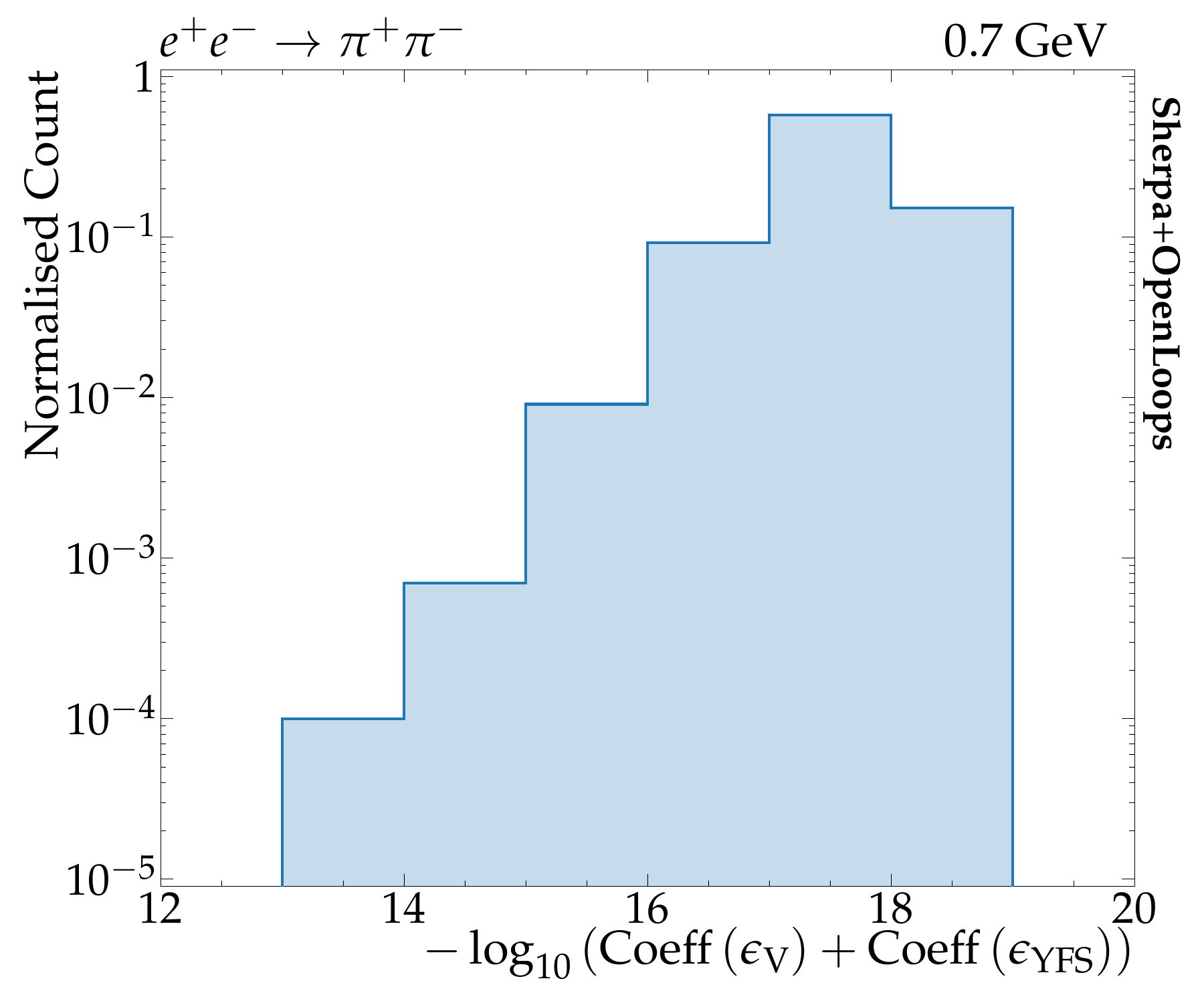}
    \includegraphics[width=0.42\linewidth]{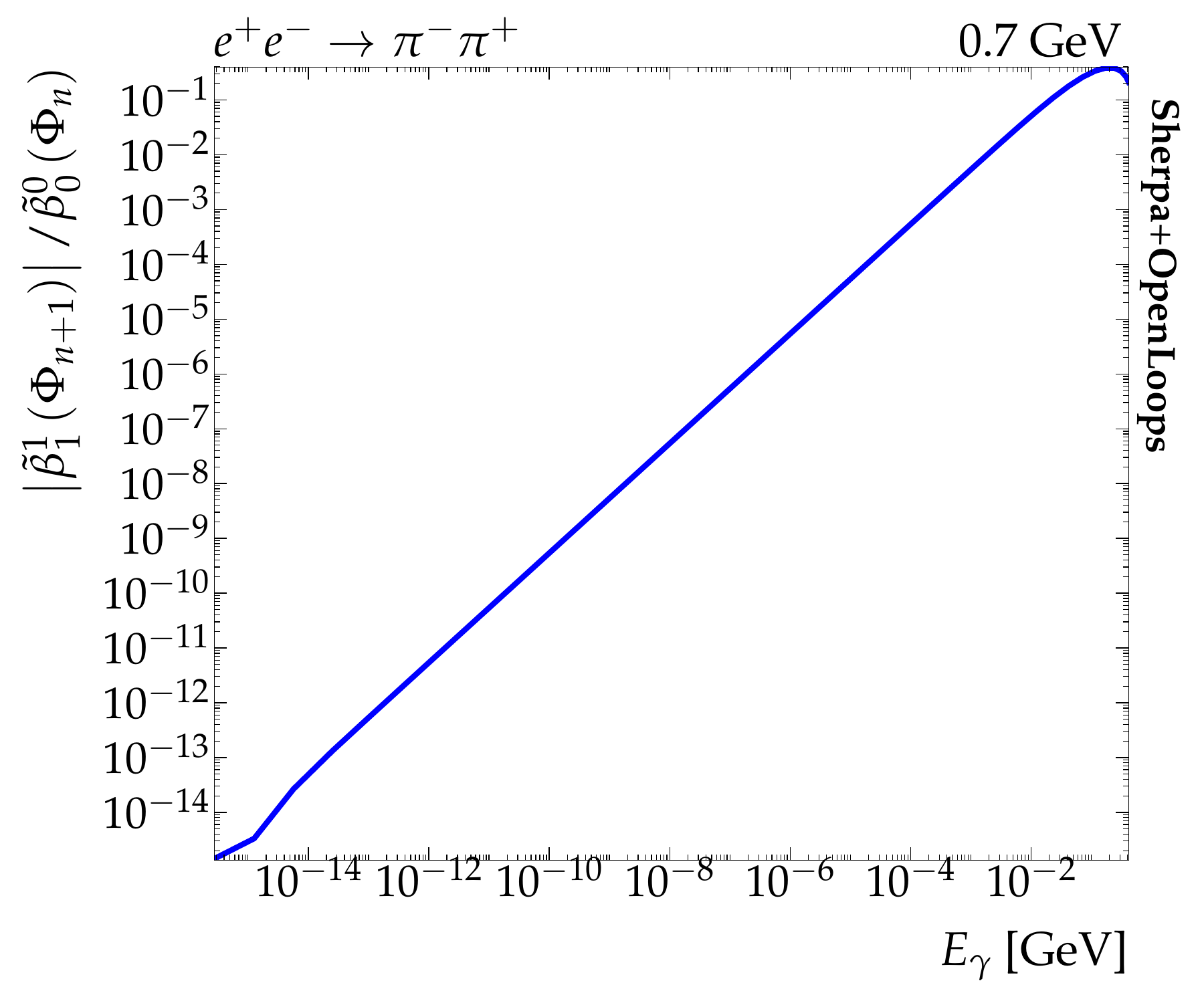}
    \includegraphics[width=0.42\linewidth]{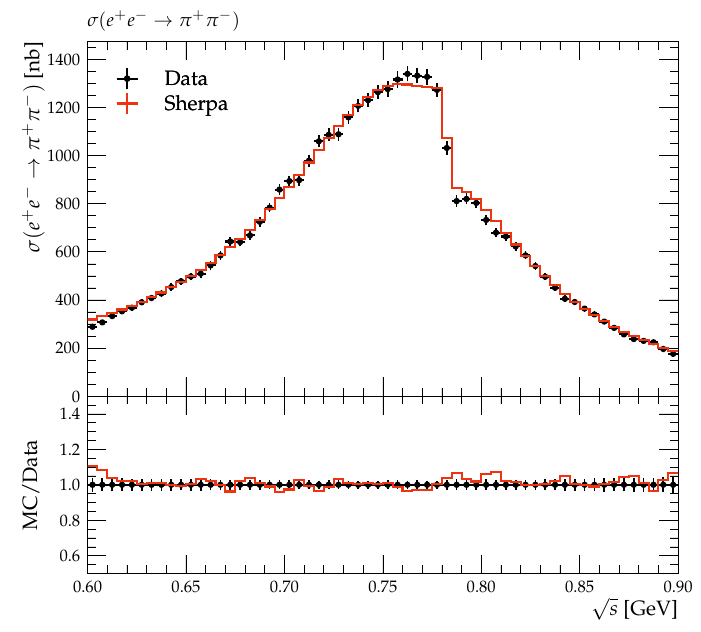}
    \caption{Left: Pole cancellation for the one-loop correction to $\ee \rightarrow  \pi^+\pi^-$ in scalar QED. Right: Scaling behaviour for the real correction to  $\pi^+\pi^-$ production. Bottom: Sherpa's prediction for the $\pi^+\pi^-$  cross-section compared with BESIII data. }
    \label{fig:Pions}
\end{figure}
\noindent
Earlier in this work we presented results that focus on future high-energy experiments, such as the future circular collider electron-positron (FCC-ee) machine, however we are not limited to such experiments. 
Indeed, due to the general purpose nature of \Sherpa we are able to apply our corrections to a wide variety of lepton experiments. 
Here we present a short list of applications, some of which will appear in separate studies.

\begin{itemize}
    \item \textbf{Low Energy \ee Experiments}: 
    Our results can be immediately applied to past and current low energy experiments~\citeLowEE, both for energy scan and radiative return experiments, for the relevant SM processes, such as Bhabha scattering and difermion production, and they can be extended to include other processes of interest such as $\pi^+\pi^-$ production.
    Indeed, the YFS approach is particularly useful here as soft photons are blind to spin, meaning that for the production of scalars, like  $\pi^+\pi^-$, we do not have to modify our resummation. 
    Of course, while the resummation remains untouched the finite \betatilde{}{} will need to include the correct amplitudes. 
    We show in~\cref{fig:Pions} how our subtraction schemes successfully removes the IR poles for pion production, enabling \Sherpa to produce \nloew predictions~\cite{Budassi:2024whw,Colangelo:2022lzg,Aliberti:2024fpq}.
    This combined with \Sherpa's native UFO~\cite{Darme:2023jdn,Hoche:2014kca} interface will allow for more accurate for beyond the standard model (BSM) modelling at \ee experiments. 
    In addition to the perturbative improvements we have included an interface to the \texttt{alphaQED}~\cite{Jegerlehner:2001ca,Jegerlehner:2006ju,Jegerlehner:2011mw} package which can be used to provide the hadronic contribution to the vacuum polarisation (HVP). 
    Furthermore, we implemented a model of the pion-form factor based on the vector meson dominance model~\cite{Sakurai:1972wk}, and our default parametrization is chosen to agree with the RadioMonteCarLow study~\cite{Aliberti:2024fpq}. 
    In the bottom plot of~\cref{fig:Pions} we compare this parametrization with Beijing Spectrometer III (BESIII) data~\cite{BESIII:2015equ}, where we have not included any additional corrections as the 
    data has undergone a dressing procedure to account for QED radiation.

    \item \textbf{Fixed Target Experiments}:  
    \Sherpa can also be used for fixed target experiments, like~\Muone~\cite{MUonE:2016hru,Gurgone:2024xdt,Banerjee:2020tdt,CarloniCalame:2015obs,CarloniCalame:2020yoz,Banerjee:2020rww} and \Moller~\cite{MOLLER:2014iki}.
    The former aims to extract HVP from scattering muons on a fixed target to produce \emupm, which \Sherpa can model using the YFS approach~\cite{Price:2025pru}.
    The latter will scatter longitudinal polarized electrons on a fixed target to induce M{\o}ller scattering, with the aim to measure the weak mixing angle, which can be modelled with \Sherpa, including the polarization of the incoming electron beam.

\end{itemize}

\section{Conclusion}

\noindent
In this work we have presented the first fully automated matching based on the YFS theorem for lepton colliders, with complete control over the \NLOEW terms and automated double-real and real-virtual corrections which arise in the \NNLOEW predications. 
We emphasize that the inclusion of double-virtual corrections is possible when such calculations become more automated and we stress that we have provided the framework in which the IR subtraction of such amplitudes can be performed automatically. 
These corrections have been implemented in the \Sherpa generator, allowing us to provide highly accurate predictions for phenomenological studies at lepton colliders. 
We have shown that our automated subtraction scheme works across a variety of processes with different final state multiplicities, by including example results from our interface to the \Griffin package.

\noindent
We also showed that our subtraction method preserves the resonance structure of the process in question at the level of the amplitude squared, without the need to resort to the CEEX approach of an amplitude-level subtraction. 
While we find good agreement with the \csew subtraction, the \yfsnlo approach significantly reduces the number of negative weights we have in the event record. 
This reduction combined with other high-performance improvements in \Sherpa~\cite{Bothmann:2022thx,Campbell:2021vlt,Bothmann:2023siu,Bothmann:2020ywa,Gao:2020zvv,Bothmann:2023gew} allows to tentatively assume that there will not be a computational bottleneck for \Sherpa at future lepton colliders for high precision simulations. We anticipate that 
the features described in this work will be made public in an upcoming release of the \Sherpa generator.

\noindent

\section*{Acknowledgements}
We would like to thank Jonas Lindert for his support with \OpenLoops, Lisong Chen and Ayres Freitas for their development and continued support of the \Griffin package, and to our fellow \Sherpa authors for their continued support. A.P would like to thank the KKMC authors and Wiesław Płaczek for many discussion on YFS and Stéphane Delorme and Januz Gluza for discussions on the radiative corrections to pions.  
The work of A.P.\ is supported by grant No. 2023/50/A/ST2/00224 of the National
Science Centre (NCN), Poland.
A.P.\ gratefully acknowledges the Polish high-performance computing infrastructure PLGrid (HPC Centers: CI TASK, ACK Cyfronet AGH) for providing computer facilities and support within computational grant no. PLG/2024/017287.
F.K.'s work is supported by STFC under grant agreement ST/P006744/1.
\bibliography{main.bib}

\end{document}